\documentclass[sigplan,10pt,nonacm]{acmart}
\renewcommand\footnotetextcopyrightpermission[1]{}
\AtBeginDocument{%
  }

\setcopyright{none} 

\usepackage{algorithm}
\usepackage{algorithmic}
\usepackage{lipsum}
\usepackage{enumitem}
\usepackage{graphicx}
\usepackage{textcomp}
\usepackage[svgnames]{xcolor}
\definecolor{saddlebrown}{RGB}{139,69,19}
\definecolor{reviewbrown}{RGB}{170,85,20}
\usepackage{listings}
\usepackage{tikz}
\usepackage{pgf-pie}
\usepackage{booktabs}   
\usepackage{tabularx}   
\usepackage{array}      
\usepackage{xspace}
\usepackage{caption}    
\usepackage[normalem]{ulem}
\usepackage{multirow}
\usetikzlibrary{positioning, arrows.meta}
\definecolor{rvvblue}{RGB}{40, 100, 180}
\definecolor{rvvgreen}{RGB}{30, 150, 80}
\definecolor{rvvorange}{RGB}{220, 100, 40}
\usepackage{float}
\usepackage{makecell}

\newcommand{\projectname}{{RVANNS}\xspace}

\usepackage[normalem]{ulem}

\begin{document}

\title{\projectname: Mixed-Precision Indexing and Locality-Aware Graph Traversal on RISC-V}

\begin{abstract}

Approximate nearest neighbor search (ANNS) on CPUs is increasingly constrained by candidate-vector movement and decoding rather than peak arithmetic throughput.
Although the RISC-V Vector Extension (RVV) provides vector-length-agnostic execution and LMUL-based register grouping, generic low-precision decoding still incurs conversion overhead, while irregular graph traversal generates scattered accesses that degrade cache locality and memory-level parallelism.

We present \projectname{},
an RVV-oriented ANNS engine that jointly optimizes vector representation and graph locality.
Its Mixed-Precision Multi-Layer Index (MPMI) represents each vector with a dense 8-bit affine base and sparse FP16/FP32 residuals, fusing reconstruction with distance accumulation and aligning widening with LMUL-sized register groups.
ROrder co-locates likely co-visited graph nodes and sorts remapped adjacency lists, transforming scattered payload probes into denser, predominantly forward-moving address streams.

Integrated into Milvus, \projectname{} achieves 3.39$\times$ and 4.94$\times$ speedups over scalar execution on real 128-bit and 256-bit RVV processors, respectively.
Under controlled HNSW configurations, it improves throughput by 2.27--2.76$\times$ over RVV SIMD+FP32 and by 1.18--1.59$\times$ over the corresponding AVX-512 and SVE baselines.
On Cohere10M, it further delivers 1.82--2.27$\times$ higher QPS/W than the evaluated GPU baselines.

\end{abstract}

\author{
Chengying Huan$^{1}$,
Yudong Liu$^{2}$,
Jianguo Wang$^{3}$,
Lizheng Chen$^{1}$,
Renling Yin$^{4}$,
Weijia Chen$^{5}$,\\
Ji Qi$^{2}$,
Jiageng Yu$^{2}$,
Junjie Xu$^{1}$,
Jie Zhang$^{6}$,
Chen Tian$^{1}$,
Yanjun Wu$^{2}$\\[4pt]
$^{1}$Nanjing University \quad
$^{2}$Institute of Software, Chinese Academy of Sciences \\
$^{3}$Purdue University \quad
$^{4}$Sichuan University \quad
$^{5}$Putian University \quad
$^{6}$Peking University
}

\renewcommand{\shortauthors}{Huan et al.}

\maketitle
\pagestyle{plain}

\section{Introduction}
Approximate nearest neighbor search (ANNS) is a core primitive in modern ML and IR systems, supporting large-scale recommendation~\cite{covington2016deep}, image retrieval and deduplication~\cite{jegou2010product}, multimodal retrieval~\cite{radford2021learning}, and semantic search for QA and RAG pipelines~\cite{karpukhin2020dense}. Given a query vector, ANNS returns its top-$k$ nearest neighbors from millions to billions of high-dimensional embeddings, avoiding the prohibitive $O(ND)$ cost of brute-force search~\cite{babenko2016efficient}, where $N$ is the dataset size and $D$ is the vector dimensionality~\cite{shrivastava2014asymmetric}. In practice, production systems such as FAISS~\cite{johnson2019billion}, Milvus~\cite{wang2021milvus}, and ScaNN~\cite{guo2020accelerating} combine hierarchical indexing (e.g., IVF) and graph-based search (e.g., HNSW~\cite{malkov2018efficient}) with SIMD- and GPU-accelerated distance kernels~\cite{fog2016instruction,johnson2017faiss} to meet latency and throughput requirements.

Although GPUs effectively accelerate ANNS, their limited on-device memory relative to hundred-gigabyte datasets often shifts the bottleneck to host--device data movement. While recent heterogeneous systems such as FusionANNS~\cite{tian2024fusionanns} and NeuPIMs~\cite{heo2024neupims} tackle this via CPU/GPU/SSD cooperation, CPU-based deployment remains a common, cost-effective choice in production. Our focus is complementary: rather than hierarchical offloading, we optimize the CPU-side ANNS backend for the RISC-V Vector Extension (RVV) through decode-path and memory-layout co-design.

A key challenge is that efficient CPU ANNS must overcome two distinct sources of inefficiency. The first is \textbf{execution-side inefficiency}: existing SIMD backends often waste useful work on tail handling, predicate management, and other control overheads, especially when the embedding dimension is not well aligned with the hardware vector shape. The second is \textbf{memory-side inefficiency}: during graph traversal and reranking, the processor repeatedly fetches vector payloads, performs decode or reconstruction, and accumulates distances over irregular candidate sets. These accesses are scattered and latency-sensitive, making ANNS an \emph{irregular, memory-bound} workload. Thus, wider SIMD units alone do not guarantee higher performance; efficient ANNS requires both an execution model that minimizes SIMD-side waste and a data path that the memory hierarchy can serve efficiently.

Despite steady progress, CPU ANNS backends remain fragmented across vendor-specific SIMD ecosystems. Fixed-width ISAs such as AVX2, AVX-512, and NEON suffer from utilization fragility when the embedding dimension $D$ is not aligned with the lane count, requiring peel loops, padding, or masked tails. Predicate-centric VLA designs, which use explicit per-lane predicate masks to control active vector elements, such as Arm SVE, reduce this fragility, but they still incur per-iteration predicate generation and propagation overhead, increasing instruction count and control work in memory-bound loops~\cite{stephens2017arm}. 

By contrast, RVV provides a particularly attractive substrate for ANNS: its state-based active-length control (\texttt{vsetvl}) sets the active vector length once per strip-mined iteration, and {\texttt{LMUL}, the vector-register group multiplier, groups multiple vector registers into one logical vector operand. For example, \texttt{LMUL}{=}4 uses four vector registers for one operand, increasing the useful work covered by each vector instruction and providing the register grouping needed by widening decode and accumulation.} These features directly reduce much of this \textbf{execution-side inefficiency}, better preserving useful memory-level parallelism (MLP).

However, porting ANNS kernels to RVV is not sufficient. Once RVV removes much of the execution-side waste, the remaining bottleneck shifts to the \textbf{memory side}. Microarchitectural profiling on real RISC-V hardware (SG2044) reveals two dominant sources of RVV efficiency loss. First, conventional decode squanders this register-grouping advantage: even at \texttt{LMUL}{=}4, an 8-bit$\rightarrow$FP32 decode path incurs \textbf{48\%} more instructions and a \textbf{47\%} cycle penalty relative to the ideal FP32 baseline, because decode-side reshaping and bookkeeping offset grouped widening and accumulation. Second, graph traversal induces scattered accesses that prevent RVV loads from exploiting locality, prefetching, and unit-stride memory behavior: irregular access nearly triples L2D MPKI (\textbf{5.15$\rightarrow$14.45}) and reduces the cycle benefit of \texttt{LMUL}{=}4 from \textbf{28.4\%} under sequential access to only \textbf{1.2\%}. These observations motivate two critical design requirements for efficient ANNS on RVV: \emph{register-group-aligned widening}, i.e., organizing low-precision decode so widened operands fill RVV \texttt{LMUL} register groups without extra reshaping, and \emph{address-monotone memory access}, i.e., arranging graph payload accesses to move mostly forward in physical address order.

To address these obstacles, we present \textbf{\projectname}, to the best of our knowledge the first end-to-end ANNS engine co-designed for RVV, jointly optimizing vector representation, graph layout, and vector execution. Although MPMI and ROrder are designed to satisfy these RVV-oriented requirements, they target generic memory-side inefficiencies in SIMD ANNS---excess bytes moved during decode and poor traversal locality---and therefore remain partially beneficial on other SIMD backends. Their strongest gains arise on RVV because \texttt{LMUL}-aligned widening and VLA execution allow reduced memory traffic and more regular access streams to translate directly into sustained end-to-end throughput gains.

The key contributions of this paper are as follows:
\begin{itemize}[leftmargin=*, topsep=0pt]
  \item \textbf{Architectural Diagnosis and Hardware-Driven Requirements.}
  We show that simply porting ANNS to RVV does not solve the underlying memory wall. Through microarchitectural profiling, we reveal that instruction bloat during generic mixed-precision decoding and severe cache thrashing during irregular graph traversal fundamentally limit RVV's scaling efficiency. This diagnosis yields two concrete, hardware-driven design requirements for RVV-efficient ANNS: \emph{register-group-aligned widening} and \emph{address-monotone memory access}.
 
  \item \textbf{MPMI for Register-Group-Aligned Widening.}
  MPMI (Mixed-Precision Multi-Layer Index) restructures mixed-precision vector representation around RVV grouped widening. By encoding vectors as a dense low-precision base with compact high-precision residuals, MPMI reduces bytes moved per candidate and fuses reconstruction with distance accumulation. This organization lets RVV's \texttt{LMUL}-based widening perform useful work without being diluted by generic FP32 materialization and reshaping overhead.

  \item \textbf{ROrder for Address-Monotone Memory Access.}
  ROrder reshapes graph traversal into a more locality-friendly memory stream. Rather than treating graph layout as an independent software preprocessing step, it co-locates frequently co-visited nodes and reorders adjacency lists by the new layout. The resulting forward-moving payload accesses help the RVV backend sustain memory-level parallelism under irregular graph traversal.
  
  \item \textbf{Architecture-Guided Realization and Evaluation.}
  We implement this co-designed framework in a production vector database and evaluate it on \textbf{real RISC-V hardware}. On SG2044 (RVV 1.0, VLEN=128\,b), \projectname achieves up to \textbf{3.39$\times$} speedup over scalar execution. {In cross-ISA ablation, \projectname improves throughput by \textbf{1.18--2.76$\times$} over each platform's SIMD baseline, with the largest gains on SG2044/RVV (\textbf{2.27--2.76$\times$}). In GPU comparison on Cohere10M, \projectname achieves \textbf{2.27$\times$} higher QPS/W than cuVS at Recall@100$\approx$0.99.}

 \end{itemize}


\section{Background and Motivation}
\subsection{ANNS Execution Characteristics}
\label{sec:background}

Approximate nearest neighbor search (ANNS)~\cite{zeng2023dfgas,liu2024juno,li2025ansmet} answers a query by pruning the search space to a candidate set and reranking those candidates, rather than exhaustively scanning the full database~\cite{indyk1998approximate,beygelzimer2006cover,cha2002comprehensive,malkov2018efficient}. In modern vector databases, this query path is typically organized as a two-stage pipeline: \emph{candidate generation} narrows the search space, and \emph{reranking} computes exact or refined distances over the surviving candidates~\cite{guo2020accelerating}. Although pruning greatly reduces the number of vectors examined, reranking still dominates end-to-end latency because it repeatedly loads vector payloads, performs dequantization when compression is used, and accumulates distances across the candidate set~\cite{li2020graf,jayaram2019diskann,ge2013optimized}. As a result, ANNS on CPUs is governed primarily by data movement rather than by peak arithmetic throughput.

\noindent\textbf{Irregular Access as a Core Execution Difficulty.}
Mainstream ANNS indexes expose different search structures but converge to the same execution difficulty. Partition-based methods such as IVF scan a few selected inverted lists, benefiting from relatively contiguous accesses. In contrast, graph-based methods such as HNSW achieve strong recall--latency trade-offs through pointer-chasing traversal over irregular neighbor lists~\cite{malkov2018efficient}. This pattern repeatedly touches scattered adjacency and payload regions, making cache behavior and prefetch timeliness central to performance.

\noindent\textbf{Compression Trades Bandwidth for Decode Overhead.}
To reduce footprint and bandwidth pressure, production systems adopt compressed representations such as SQ and PQ~\cite{gong2012iterative,wei2020pq}. Compression reduces per-candidate bytes moved, but introduces decode overhead during reranking via reconstruction or asymmetric distance computation. Therefore, for memory-bound ANNS, the key objective is to reduce bytes moved without introducing decode-side execution overhead that negates the bandwidth savings.

\vspace{-2mm}

\subsection{SIMD Execution Models for ANNS}
\label{sec:bg-rvv}

SIMD backends for ANNS differ mainly in handling variable dimensions and the execution-side overhead introduced during memory stalls. Fixed-width ISAs such as AVX2 and NEON execute a constant number of elements per instruction; when the embedding dimension $D$ is not a multiple of the lane count, implementations must introduce peel loops, padding, or masked tails. This wastes lanes on unaligned dimensions and increases backend specialization. Table~\ref{tab:simd_comparison} summarizes the ISA features most relevant to ANNS, including vector-length semantics, tail handling, predication, register scaling, and indexed memory support.

\begin{table}[t]
\centering
\caption{Comparison of current SIMD ISAs relevant to ANNS.}
\label{tab:simd_comparison}
\footnotesize
\setlength{\tabcolsep}{1.2pt}
\begin{tabular}{lccccc}
\toprule
\textbf{Feature} & \textbf{AVX2} & \textbf{AVX-512} & \textbf{SVE} & \textbf{NEON} & \textbf{RVV} \\
\midrule
Vector length & 256b fixed & 512b fixed & VLA (mask) & 128b fixed & VLA (\texttt{vsetvl}) \\
Tail control & Peel loops & Masking & \texttt{pg}+\texttt{svwhilelt} & Peel loops & \texttt{vsetvl} state \\
Predication & Partial & Full & Full & Partial & Full \\
Register scaling & No & No & No & No & \texttt{LMUL} \\
Indexed access & Limited & Better & Yes & Limited & Yes \\
\bottomrule
\end{tabular}
\end{table}

\noindent\textbf{The Right Comparison Is MLP, Not Peak ALU Speed.}
For memory-bound ANNS, the most important question is not which ISA offers the highest nominal vector throughput, but which execution model better preserves useful memory-level parallelism (MLP) under irregular access. Fixed-width SIMD already loses efficiency to tail handling. Predicate-centric VLA designs such as Arm SVE remove that fragility, but they introduce a different overhead: strip-mined loops repeatedly generate predicates (e.g., \texttt{svwhilelt}) and propagate predicate state through subsequent vector instructions. In irregular ANNS kernels, this predicate-management overhead increases instruction count and control work, reducing the effective in-flight capacity available to hide memory latency.

\noindent\textbf{RVV as a Promising Execution Substrate.}
The RISC-V Vector Extension (RVV) adopts a state-based VLA model: \texttt{vsetvl} configures the active vector length once per strip-mined iteration and exposes it as architectural state (\texttt{vl}) for subsequent vector instructions~\cite{rvv2021spec,riscv-unprivileged,zaruba2019cost,platzer2021ten}. RVV also provides indexed memory operations for irregular data structures and {\texttt{LMUL} (the vector-register group multiplier), which groups physical vector registers into one logical vector operand for widening and accumulation.} Compared with fixed-width SIMD and predicate-centric VLA, RVV removes more execution-side inefficiencies and better preserves useful MLP under ANNS workloads. It is therefore a promising substrate for ANNS, but not yet a complete solution.

\noindent\textbf{Compression Choice for SIMD ANNS: SQ over PQ.}
As noted in FLASH~\cite{flash2025sigmod}, SQ is generally more suitable than product quantization (PQ) for CPU SIMD backends. PQ reconstructs distances via indirect, gather-dominated codebook lookups that weaken spatial locality and hinder efficient vector execution. By contrast, SQ uses a dense, sequential representation with shared scaling metadata. This maps naturally to widening loads and lightweight reconstruction, making it more cache-friendly and less irregular on the decode path.

\noindent\textbf{Key Terms.}
The discussion above introduces several RVV/SIMD terms used throughout the paper.
Table~\ref{tab:key_terms} briefly summarizes the execution concepts needed for the following motivation, design, and evaluation sections.

\begin{table}[t]
\centering
{
\caption{{Key RVV/SIMD execution terms.}}
\label{tab:key_terms}
\scriptsize
\setlength{\tabcolsep}{3pt}
\begin{tabular}{p{0.36\columnwidth}p{0.56\columnwidth}}
\toprule
\textbf{Term} & \textbf{Brief Explanation} \\
\midrule
RVV & RISC-V vector extension for CPU SIMD execution. \\
VLA & Vector-length agnostic; vector length is set at runtime. \\
\texttt{LMUL} & Groups RVV registers into wider logical operands. \\
\texttt{vsetvl}/\texttt{vl} & Sets the active vector length in RVV. \\
MPKI & Cache misses per thousand instructions. \\
Predicate-centric VLA & SVE-style execution with explicit lane masks. \\
Register-group-aligned widening & Decode aligned with \texttt{LMUL}-sized FP32 operands. \\
\bottomrule
\end{tabular}
}
\end{table}

\begin{figure}[t]
  \centering
  \includegraphics[width=\linewidth]{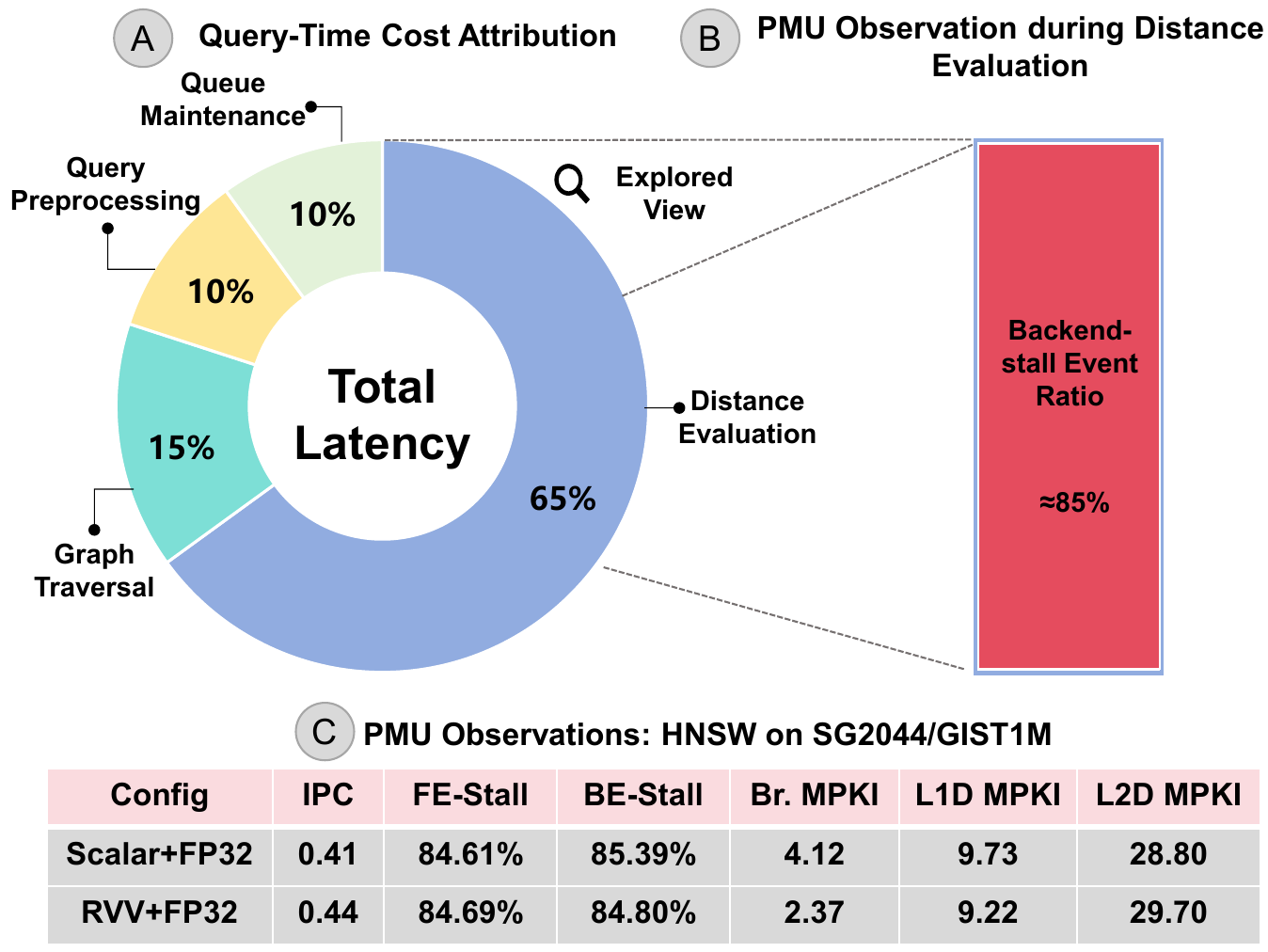}
    \caption{{ANNS query latency and stall profile on SG2044. (A) Query-phase breakdown of graph-based ANNS. (B) PMU-based stalled-cycle breakdown of distance evaluation. (C) PMU counters with HNSW on GIST1M. IPC denotes instructions per cycle; FE/BE stall and MPKI report frontend/backend stalls and cache/branch misses.}}
  \label{fig:anns_time_distribution}
\end{figure}

\vspace{-2mm}

\subsection{Motivation}
\label{sec:motivation}
\noindent\textbf{The Generic Bottleneck: SIMD-Accelerated ANNS Is Memory-Bound.}
Before focusing on RVV, we first identify the bottleneck of ANNS on CPUs. Prior ANNS systems and graph-reordering studies report that graph-based near-neighbor search is often limited by irregular memory access and cache misses~\cite{ding2019quicker,jayaram2019diskann,coleman2021graphreorder}. As illustrated in Fig.~\ref{fig:anns_time_distribution}(A), the \emph{distance evaluation} phase dominates end-to-end query latency because it repeatedly loads vector payloads, performs reconstruction when compression is used, and accumulates distances over visited candidates.

{Fig.~\ref{fig:anns_time_distribution}(C) reports PMU profile for HNSW on GIST1M. Scalar+FP32 and RVV+FP32 show low IPC (0.41/0.44) and high frontend/backend stalls, while L2D MPKI remains high (28.80/29.70) and branch-miss MPKI is smaller (4.12/2.37). Thus, vectorization changes execution form but leaves the irregular payload-access bottleneck intact, indicating that graph-based SIMD ANNS is limited by memory-side pressure rather than speculation or peak arithmetic throughput.}

\noindent\textbf{{Architectural Insight: RVV Removes Execution-side Waste, Thereby Exposing the Memory Wall More Clearly.}}
{The memory-side pressure above is generic to graph-based ANNS, as it follows from irregular candidate traversal and payload fetches rather than from a particular SIMD ISA. Fig.~\ref{fig:anns_time_distribution} provides the direct PMU evidence for this effect on SG2044; later cross-ISA ablations evaluate how the same memory-side mechanisms transfer to SVE and AVX-512.} The execution substrate nevertheless dictates how clearly this limit manifests. Fixed-width and predicate-centric ISAs often obscure this memory limit behind execution-side inefficiencies such as explicit tail handling and predicate generation. By directly reducing these overheads through state-based control (\texttt{vsetvl}) and register grouping (\texttt{LMUL}), RVV makes the remaining memory limit more visible. Because RVV does not inherently increase physical memory bandwidth, ANNS performance on RVV depends more directly on whether the memory system can feed the vector pipeline efficiently.

\begin{table}[t]
\centering
\caption{RVV decode-path profiling on SG2044. Controlled \texttt{LMUL=1}/\texttt{LMUL=4} runs isolate register-grouping benefits and mixed-precision decode overhead.}
\label{tab:lmul_profiling}
\footnotesize
\setlength{\tabcolsep}{3pt}
\begin{tabular}{lccccc}
\toprule
\textbf{Kernel} & \textbf{Target \texttt{LMUL}} & \textbf{Cycles (M)} & \textbf{Instr. (M)} & \textbf{IPC} \\ \midrule
Ideal FP32 Baseline & \texttt{LMUL} = 1 & 117.1 & 241.9 & 2.07 \\
Ideal FP32 Baseline & \texttt{LMUL} = 4 & 56.6  & 62.0  & 1.10 \\ 
Conventional SQ Decode  & \texttt{LMUL} = 1 & 484.4 & 362.0 & 0.75 \\
Conventional SQ Decode  & \texttt{LMUL} = 4 & 83.6  & 92.0  & 1.10 \\ \bottomrule
\end{tabular}
\end{table}

\noindent\textbf{Challenge 1: Conventional Decode Dilutes RVV's Widening Advantage.}
While low-precision storage reduces memory traffic, conventional SQ decode can offset this benefit with execution-side overhead. Here, \emph{conventional SQ decode} denotes a generic 8-bit$\rightarrow$FP32 reconstruction path without RVV-oriented packing and fusion, i.e., without register-group-aligned widening. Table~\ref{tab:lmul_profiling} shows that RVV register grouping is effective for FP32: moving from \texttt{LMUL=1} to \texttt{LMUL=4} reduces cycles from 117.1M to 56.6M. However, conventional SQ decode exposes two penalties:

\begin{itemize}[leftmargin=*, topsep=2pt, itemsep=2pt]
\item \textbf{Fractional \texttt{LMUL} Penalty:} Decoding 8-bit data to a \texttt{LMUL=1} FP32 target forces fractional-\texttt{LMUL} loads (\texttt{LMUL=1/4}), reducing IPC to 0.75 and increasing execution to 484.4M cycles.
\item \textbf{Instruction Bloat:} \texttt{LMUL=4} avoids the fractional path but still requires multi-stage widening and conversion; it executes 48\% more instructions than the ideal FP32 \texttt{LMUL=4} baseline (92.0M vs. 62.0M), incurring a 47\% cycle penalty.
\end{itemize}

Consequently, conventional decode turns memory savings into an instruction-heavy path that cannot fully exploit RVV register grouping. We therefore quantify the penalty with cycles and instruction counts rather than IPC alone, and require a representation that preserves register-group-aligned widening.

\begin{table}[t]
\centering
\caption{Profiling RVV distance evaluation on SG2044. Irregular traversal causes L2D miss pressure, neutralizing the cycle benefits of \texttt{LMUL}-grouped execution.}
\label{tab:challenge2_profiling}
\resizebox{0.48\textwidth}{!}{
\begin{tabular}{lccccc}
\toprule
\textbf{Access Pattern} & \textbf{\texttt{LMUL}} & \textbf{Cycles (M)} & \textbf{Instr. (M)} & \textbf{L1D MPKI} & \textbf{L2D MPKI} \\ \midrule
Sequential & \texttt{LMUL} = 1 & 344.3 & 98.7  & 4.95 & 5.15 \\
Sequential & \texttt{LMUL} = 4 & 246.4 & 84.4  & 5.80 & 6.07 \\ 
Irregular  & \texttt{LMUL} = 1 & 289.7 & 101.7 & 4.27 & 14.45 \\
Irregular  & \texttt{LMUL} = 4 & 286.3 & 87.3  & 6.33 & 16.02 \\ \bottomrule
\end{tabular}
}
\end{table}

\noindent\textbf{Challenge 2: Irregular Traversal Neutralizes RVV's \texttt{LMUL} Scaling.}
Even with an efficient decode path, graph-based ANNS fetches payloads in traversal order rather than physical address order. In Table~\ref{tab:challenge2_profiling}, \emph{irregular} denotes pointer-chasing payload accesses over the same layout. Such scattered accesses limit locality and impose two penalties on RVV execution:

\begin{itemize}[leftmargin=*, topsep=2pt, itemsep=2pt]
\item \textbf{Cache Thrashing:} Irregular traversal nearly triples L2D MPKI compared with sequential access (5.15 $\rightarrow$ 14.45), reducing useful memory-level parallelism.
\item \textbf{Neutralized \texttt{LMUL} Scaling:} Sequential access benefits from the \texttt{LMUL=4} grouping (344.3M $\rightarrow$ 246.4M cycles), whereas under irregular access the gain almost vanishes (289.7M $\rightarrow$ 286.3M).
\end{itemize}

Consequently, irregular payload access leaves RVV dominated by memory service rather than vector execution, preventing grouped vector instructions from yielding cycle reduction. This motivates a traversal-aware layout that improves address monotonicity~\cite{wang2024ndsearch}.




\noindent\textbf{Design Principles for Efficient ANNS on RVV.}
These observations indicate that RVV's benefit on ANNS does not come from vectorization alone. To translate RVV's execution advantages into end-to-end performance, the ANNS data path should preserve grouped widening during decode and provide locality-friendly payload streams during traversal.

\noindent\textbf{{Register-Group-Aligned Widening.}}
The decode path should organize low-precision data so that widening and FP32 accumulation align with \texttt{LMUL}-based register groups, avoiding the reshaping overhead of materializing a reconstructed FP32 vector.

\noindent\textbf{{Address-Monotone Memory Access.}}
Traversal-induced payload accesses should be reshaped into locality-friendly, forward-moving address streams, allowing RVV loads to exploit spatial locality, prefetching, and sustained MLP.


\section{\projectname Engine Framework}
\label{sec:framework}

\begin{figure*}[t]
    \centering
    \includegraphics[width=.95\textwidth]{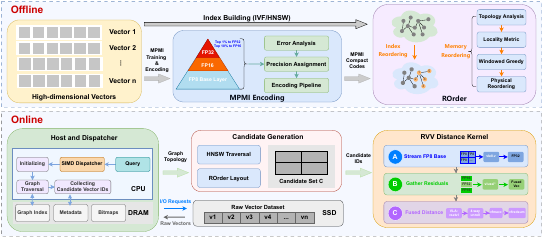}
    \caption{
     System architecture of \projectname. The offline phase constructs MPMI for register-group-aligned widening and applies ROrder for address-monotone memory access. During online search, the shared RVV kernel consumes the reordered graph and mixed-precision payloads through fused decode-and-distance execution, translating both into higher query throughput.
    }
    \label{fig:rvanne_arch}
\end{figure*}

\subsection{Overview}
\label{sec:arch}

Guided by the design principles in Sec.~\ref{sec:motivation}, we propose \projectname to overcome ANNS bandwidth and execution bottlenecks on RVV. As illustrated in Fig.~\ref{fig:rvanne_arch}, \projectname consists of two core mechanisms:

\begin{itemize}[leftmargin=*, topsep=0pt]
    \item \textbf{{MPMI for register-group-aligned widening (Sec.~\ref{sec:MPMI}).}}
    MPMI restructures mixed-precision vector representation so that a dense low-precision base and sparse high-precision residuals can be widened and fused with substantially less reshaping overhead. This allows RVV's \texttt{LMUL}-based widening and accumulation to scale more useful work along the decode-and-distance path.

    \item \textbf{{ROrder for address-monotone memory access (Sec.~\ref{sec:reorder}).}}
    ROrder reorganizes graph layout according to traversal co-visitation patterns, reshaping scattered accesses into more address-monotone and locality-friendly streams. This facilitates unit-stride RVV loads where possible and reduces the penalty of indexed gathers.
\end{itemize}

This framework follows a one-to-one mapping from bottlenecks to mechanisms and validations. Challenge~1 motivates the register-group-aligned widening requirement, realized by MPMI and validated in Sec.~\ref{sec:eval_mpmi}; Challenge~2 motivates the address-monotone memory access requirement, realized by ROrder and validated in Sec.~\ref{ablation_3}. The RVV kernel layer serves as the common execution substrate, whose standalone effect is isolated in Sec.~\ref{sec:generality-vlen}. {While the memory-side effects partially transfer to non-RVV SIMD backends, the mechanisms remain RVV-oriented: MPMI is guided by \texttt{LMUL}-aligned widening, and ROrder is guided by the need to preserve RVV load locality and memory-level parallelism under irregular traversal.}

\begin{algorithm}[t]
\caption{{\projectname End-to-End Workflow}}
\label{alg:RVANNS_workflow}
\small
\begin{algorithmic}[1]
{
\renewcommand{\algorithmicrequire}{\textbf{Input:}}
\renewcommand{\algorithmicensure}{\textbf{Output:}}
\REQUIRE Database $\mathcal{X}\!\subset\!\mathbb{R}^D$, queries $\mathcal{Q}$, HNSW params $(M, efC, efS)$, MPMI thresholds $(\tau_{32}, \tau_{16})$, ROrder window $w$, result size $k$
\ENSURE Top-$k$ nearest neighbors for each $q \in \mathcal{Q}$
\STATE $G, \mathit{idx} \leftarrow \textsc{MPMI\text{-}Build}(\mathcal{X}, M, efC, \tau_{32}, \tau_{16})$ \hfill // Alg.~\ref{alg:fused_decode_dist}
\STATE $\pi, G' \leftarrow \textsc{ROrder}(G, w, \mathit{idx})$ \hfill // Alg.~\ref{alg:rorder}
\FOR{each query $q \in \mathcal{Q}$}
    \STATE $C \leftarrow \textsc{BeamSearch}(G', q, efS)$ \hfill // traverse reordered graph
    \FOR{each candidate $c \in C$}
        \STATE $d_c \leftarrow \textsc{MPMI\text{-}FusedDist}(q, \mathit{idx}[c])$ \hfill // Alg.~\ref{alg:fused_decode_dist}
    \ENDFOR
    \STATE $\mathit{Results}[q] \leftarrow$ top-$k$ from $C$ ranked by $d_c$
\ENDFOR
\RETURN $\mathit{Results}$
}
\end{algorithmic}
\end{algorithm}

{\noindent\textbf{Workflow.} Alg.~\ref{alg:RVANNS_workflow} summarizes the end-to-end workflow of \projectname. 
Lines~1--2 correspond to the offline stage: MPMI-Build constructs the HNSW graph together with the MPMI compact representation, and ROrder then materializes the reordered graph layout. 
Lines~3--10 correspond to the online query stage: each query traverses the reordered graph to obtain candidates, computes candidate distances through MPMI fused decode-and-distance evaluation, and returns the top-$k$ nearest neighbors.
}

\subsection{Supporting RVV Kernel Layer}
\label{sec:rvv-distance}

\projectname uses a shared RVV kernel layer to realize the two mechanisms above. They serve as the execution substrate through which MPMI and ROrder are translated into efficient fused decode-and-distance execution.

\begin{itemize}[leftmargin=*, topsep=0pt]

    \item \textbf{VLA Execution.}
    Kernels set the active vector length once per strip-mined loop with \texttt{vsetvl} / \texttt{vsetvli}, avoiding explicit tail code and unnecessary control overhead. Distance computation then proceeds with tiled \texttt{vfsub}+\texttt{vfmacc} using dynamic \texttt{vl} to keep lanes utilized across the final tile.

    \item \textbf{Register Blocking and Unrolling.}
    Hot loops use four independent accumulators to expose ILP and hide memory latency:
    \(
    \textstyle \mathrm{acc} \gets \sum_{j=0}^{3}\texttt{vfmacc}(\mathbf{x}_{4i+j},\mathbf{y}_{4i+j},\mathrm{acc}_j),
    \)
    followed by vector reductions, including \texttt{vfadd} and \texttt{vfredsum.vs}. This overlaps loads with FMAs and sustains pipeline occupancy.

    \item \textbf{Precision-Aware Widening.}
    Mixed-precision paths widen on entry and narrow on exit: FP16/BF16 use \texttt{vfwcvt.f.f.v} before FMAs; INT8 paths use widening ops for products and accumulate in wider lanes. This minimizes conversion overhead while preserving numerical accuracy for quantized indices.

\end{itemize}

\noindent\textbf{Variable-Length Execution in Practice.}
RVV's VLA model is used in three ways. First, runtime \texttt{vl} adapts naturally to the remaining dimensions, eliminating explicit tail code; for example, on VLEN=128\,b with LMUL=4, \texttt{vl}=16 for FP32 lanes, so $D{=}961$ is handled by a final iteration with \texttt{vl}=1 and no separate tail loop. Second, the same kernel structure remains valid across different VLEN implementations, as validated in Sec.~\ref{sec:generality-vlen}. Third, MPMI decode also operates under the architectural active length, allowing sparse FP16/FP32 residual segments to be processed without dimension-specific control flow. Together, these properties let a single RVV code path handle arbitrary $D$ and different VLEN configurations while avoiding peel loops and explicit padding.

\vspace{-2mm}
\section{Mixed-Precision Multi-Layer Index}
\label{sec:MPMI}

To satisfy register-group-aligned widening, MPMI stores each vector as a dense 8-bit affine base plus sparse FP16/FP32 residuals, reducing payload traffic while enabling fused decode-and-distance execution on RVV.

\vspace{-1mm}

\subsection{Design Overview}

\begin{figure}[t]
    \centering
     \includegraphics[width=0.49\textwidth]{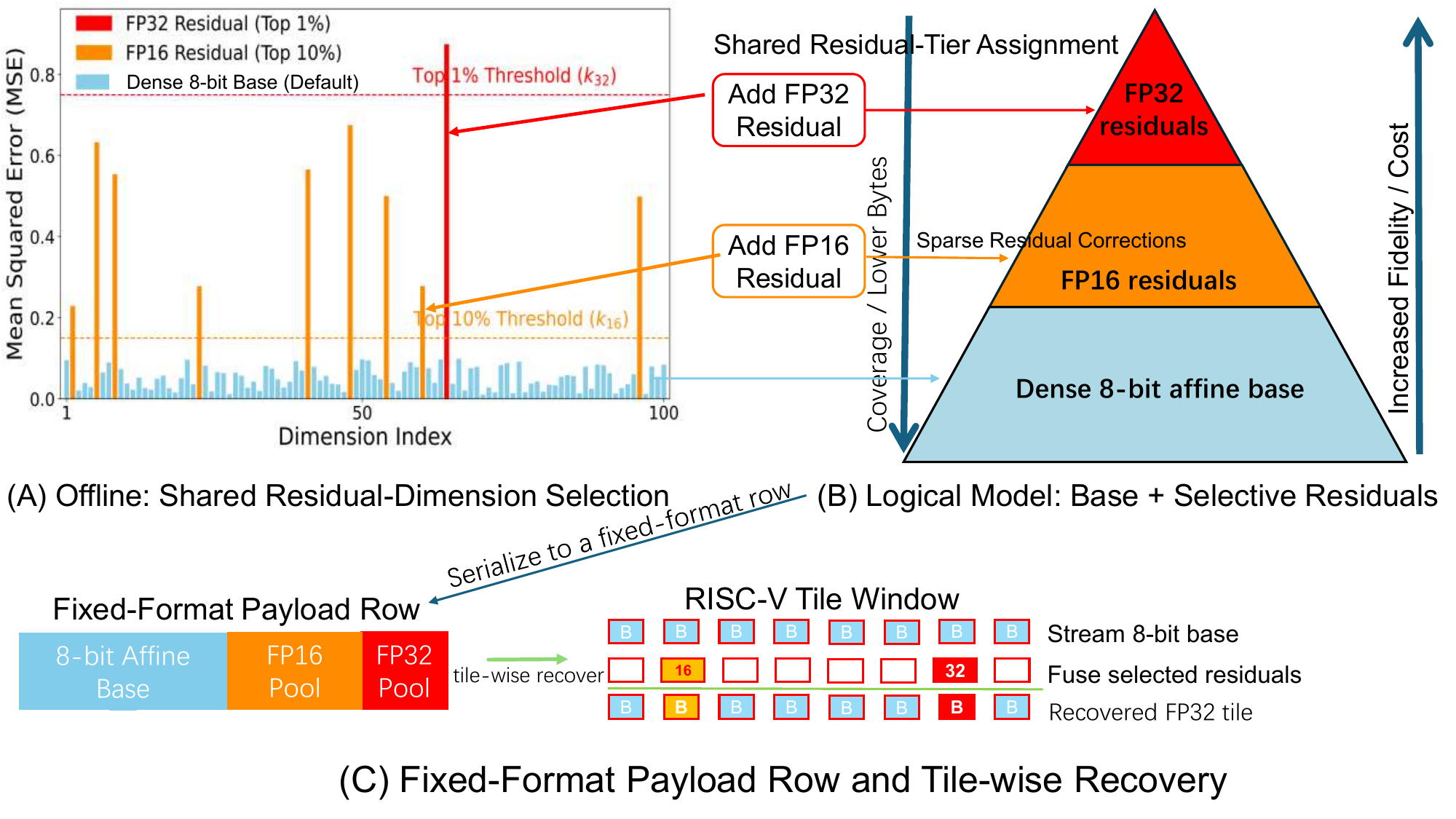}
    \caption{
        Mixed-Precision Multi-Layer Index (MPMI). (A) Sensitivity Analysis: High-error (MSE) dimensions are assigned to the FP32 or FP16 residual tier. (B) Logical Model: A cache-aware pyramid with a dense 8-bit affine base and sparse FP16/FP32 residuals. (C) Physical Layout and LMUL-Aligned RVV Decode: the contiguous 8-bit affine base is widened from an LMUL=1 base group to an LMUL=4 FP32 decode group, with sparse FP16/FP32 residual fusion on selected lanes.
    }
    \label{fig:mpmi_overview}
\end{figure}

MPMI uses a dense 8-bit affine base and compact FP16/FP32 residual pools. Selection bitmaps and prefix arrays enable on-the-fly residual fusion without materializing a full FP32 vector (Fig.~\ref{fig:mpmi_overview}(B)--(C)).

\noindent\textbf{Definition.}
MPMI encodes $x \in \mathbb{R}^D$ as $\mathcal{E}(x) = (b,\, \mathbf{r}^{16},\, \mathbf{r}^{32})$,
where $b \in \{0,\ldots,255\}^D$ is a dense 8-bit base for all dimensions, and $\mathbf{r}^{16}$ and $\mathbf{r}^{32}$ provide higher-precision residual corrections for the dimensions assigned to the FP16 and FP32 tiers, respectively. 

\noindent\textbf{Decoding Model and Query Workflow.}
Querying with MPMI avoids upfront decoding: the kernel reconstructs each vector 
on-the-fly and the result is immediately consumed by fused multiply-accumulate 
(FMA) operations, without materializing a persistent full vector in memory.
The reconstructed value of dimension $i$ is
\begin{equation}
\hat{x}_i = 
    \underbrace{\big(v_{\min}[i] + v_{\mathrm{diff}}[i] \cdot \frac{b_i}{255.0}
    \big)}_{\text{8-bit affine base (all dims)}}
\;+\; \mathbf{1}_{i\in \mathcal{S}_{16}} \cdot r^{16}_i 
\;+\; \mathbf{1}_{i\in \mathcal{S}_{32}} \cdot r^{32}_i.
\label{eq:mpmi-recon}
\end{equation}
Here $\mathbf{1}_{i\in \mathcal{S}_{16}}$ and $\mathbf{1}_{i\in \mathcal{S}_{32}}$
are indicator functions that equal 1 if dimension $i$ belongs to the
corresponding residual set, and 0 otherwise;
$r^{16}_i$ and $r^{32}_i$ store residual corrections to the 8-bit affine 
base for dimensions in $\mathcal{S}_{16}$ and 
$\mathcal{S}_{32}$ respectively.


\subsection{Compression and Quantization}
\label{sec:mpmi-compress}

The encoder treats precision assignment as a rate-distortion problem, running a one-time offline pass over database vectors in Alg.~\ref{alg:fused_decode_dist}, Lines~4--7 (no training set):
(1) compute per-dimension affine parameters $(v_{\min}[j], v_{\mathrm{diff}}[j])$,
clamping $v_{\mathrm{diff}}[j] \leftarrow \max(v_{\mathrm{diff}}[j], 10^{-10})$;
(2) quantize into a dense 8-bit affine base $b \in \{0,\ldots,255\}^D$;
(3) accumulate residual MSE scores $e_j$ in FP64;
(4) assign $\mathcal{S}_{32}$ (FP32) to the top-$\tau_{32}$ and $\mathcal{S}_{16}$ 
(FP16) to the next top-$\tau_{16}$ by $e_j$, with mutual exclusion enforcing 
FP32 priority. The \texttt{code\_size} is determined by this static 
allocation, not an input budget. Both are tunable.

\noindent\textbf{Metadata Serialization.}
Metadata is stored once per quantizer rather than per encoded vector: 
(i) FP16 and FP32 \emph{selection bitmaps}, packed into 32-bit blocks;
(ii) per-dimension $(v_{\min}, v_{\mathrm{diff}})$ scales.

During decoder initialization, \texttt{build\_prefixes\_()} derives prefix arrays from the bitmaps to map logical dimension positions to offsets in the compact residual pools in $O(1)$ time. This organization keeps each vector code as a dense 8-bit affine base followed by compact FP16/FP32 residual pools, while turning residual recovery into a bitmap- and offset-driven path compatible with RVV mask construction, population counting, and indexed loads.

\noindent\textbf{RVV-Accelerated Encoding and Layout.}
The encoding pipeline is vectorized using RVV. After computing residuals, it uses the pre-computed bitmaps to pack them. This process maps naturally to RVV's VLA model: it uses \texttt{vcpop} to determine the number of sparse residuals, sets the vector length (\texttt{vl}) accordingly via \texttt{vsetvl}, and then uses \texttt{vcompress} to pack active residuals into dense pools. This reduces the branches, peel loops, and masking overhead that would otherwise arise in fixed-width SIMD implementations. The serialized layout for each vector's code is a  compact block: a contiguous 8-bit affine base followed by the packed FP16 and FP32 residual pools, matching the layout in Fig.~3(C) and Alg.~\ref{alg:fused_decode_dist}, Lines~8--9.


\subsection{Decoding and RVV Integration}
\label{sec:decode}
During query evaluation, MPMI reconstructs vectors on the fly at vector-tile granularity. The dense 8-bit affine base is consumed as a regular stream, while FP16/FP32 corrections are recovered for bitmap-selected dimensions through population counting, prefix-based offset lookup, and indexed loads. RVV's VLA execution handles non-multiple dimensions without peel loops, so the tile loop supports arbitrary dimensionalities and sparse residual sets.

\noindent\textbf{8-bit Affine Base Reconstruction.}
The decoder streams the 8-bit affine base using wide RVV loads, converting bytes to floats via widening instructions (e.g., \texttt{vwcvtu} $\rightarrow$ \texttt{vfcvt.f.xu.v}). The base is reconstructed with a fused affine transform:
\(
x_{\text{base}} = v_{\min} + \big(b / 255.0\big) \times v_{\mathrm{diff}}.
\)
A predicate $m_{\text{nz}} = (v_{\mathrm{diff}} > 0)$ guards divisions to ensure stability without branches. RVV's VLA semantics naturally handle dimensions that are not multiples of the physical vector width, eliminating peel loops.

\noindent\textbf{Residual Fusion.}
Residuals are identified by bitmap masks (using \texttt{vmsne}, \texttt{vcpop}). A simple masked add (\texttt{vfadd\_m}) is invalid because the dense base vector (length \texttt{vl}) and the sparse residual vector (length \texttt{pop\_cnt}) are not element-wise aligned. \texttt{vluxei*} loads the selected residuals, \texttt{vrgather} packs the corresponding base lanes, and \texttt{vsuxei*} writes the corrected values into a temporary buffer. The downstream distance kernel then reloads this buffer.

\noindent\textbf{Fused Decode\texorpdfstring{$\rightarrow$}{->}Distance Kernels.}
The corrected tile feeds the L2 or inner-product distance kernel. Accumulation uses fused multiply-accumulate (\texttt{vfmacc.vv}), with modest unrolling and multiple independent accumulators to expose instruction-level parallelism. In the mixed-precision path, reconstruction and distance evaluation are organized around LMUL-friendly FP32 groups, reducing reshaping overhead while keeping metric selection resolved at compile time. {Alg.~\ref{alg:fused_decode_dist} connects these mechanisms to implementation: Lines~4--11 build the affine quantizer, residual metadata, and compact codes once offline, while Lines~17--28 perform the online RVV tile loop that streams the 8-bit base, fuses selected residuals, and accumulates distance without materializing a persistent full FP32 vector.}

\begin{algorithm}[t]
  \caption{{MPMI: Build and Fused Decode-Distance}}
  \label{alg:fused_decode_dist}
  \small
  \begin{algorithmic}[1]
  {
  \STATE \textbf{Function} \textsc{MPMI-Build}$(\mathcal{X}, M, efC, \tau_{32}, \tau_{16})$
  \STATE \quad \textbf{Input:} database vectors $\mathcal{X}$, graph params $M, efC$, thresholds $\tau_{32},\tau_{16}$
  \STATE \quad \textbf{Output:} graph $G$, MPMI index $\mathit{idx}$
  \STATE \quad $v_{\min}, v_{\mathrm{diff}} \leftarrow \textsc{TrainAffine}(\mathcal{X})$; \; $v_{\mathrm{diff}}[j]
  \leftarrow \max(v_{\mathrm{diff}}[j], \epsilon)$
  \STATE \quad $e_j \leftarrow \sum_{x \in \mathcal{X}} (x_j - \mathrm{dequant}(x_j))^2 / |\mathcal{X}|$ for each dim
  $j$
  \STATE \quad $\mathcal{S}_{32} \leftarrow \text{top-}\tau_{32}\text{ dims by } e_j$;\; $\mathcal{S}_{16} \leftarrow \text{next-}\tau_{16}$;\; 
  \STATE \quad pack bitmaps $B_{16}, B_{32}$ and prefix arrays $P_{16}, P_{32}$
  \STATE \quad \textbf{for} each $x \in \mathcal{X}$ \textbf{do}                                   
  \STATE \qquad Encode $x$ into $\texttt{code}$: $[\text{FP8 base} \| \text{FP16 residuals} \| \text{FP32 residuals}]$    
  \STATE \quad $G, \mathit{idx} \leftarrow \textsc{BuildGraph}(\mathcal{X}_{\text{enc}},\, M,\, efC)$
  \STATE \quad \textbf{return} $G, \mathit{idx}$
  \STATE
  \STATE \textbf{Function} \textsc{MPMI-FusedDist}$(q, \texttt{code})$
  \STATE \quad \textbf{Input:} query $q \in \mathbb{R}^D$, MPMI-encoded candidate code
  \STATE \quad \textbf{Output:} squared L2 distance $\|q - \hat{x}\|^2$
  \STATE \quad $\mathit{acc} \leftarrow 0$; \; $i \leftarrow 0$
  \STATE \quad \textbf{while} $i < D$ \textbf{do}
  \STATE \qquad $vl \leftarrow \texttt{vsetvl}(D - i, \texttt{e32m4})$ \hfill // VLA: set active length
  \STATE \qquad $\mathbf{vf} \leftarrow
  \texttt{vfcvt}(\texttt{vwcvtu}(\texttt{vwcvtu}(\texttt{vle8}(\texttt{code}{+}i))))$ 
  \STATE \qquad $\mathbf{base} \leftarrow v_{\min}[i{:}] + \mathbf{vf} / 255 \times v_{\mathrm{diff}}[i{:}]$ \hfill //
  affine dequant
  \STATE \qquad $m_{16} \leftarrow \texttt{vmsne}(B_{16}[i{:}], 0)$; \; $n_{16} \leftarrow \texttt{vcpop}(m_{16})$
  \STATE \qquad \textbf{if} $n_{16} > 0$ \textbf{then}
  \STATE \qquad \quad $\mathbf{r} \leftarrow \texttt{vluxei32}(\mathit{pool}_{16},\; P_{16}[i],\; n_{16})$ \hfill //
  indexed gather
  \STATE \qquad \quad $\mathbf{base} \leftarrow \texttt{vfadd}(\mathbf{base},\; \texttt{vfwcvt}(\mathbf{r}),\; m_{16})$
  \hfill // masked fuse
  \STATE \qquad FP32 residual: same with $\texttt{vluxei32}$, $B_{32}$, $P_{32}$
  \STATE \qquad $\mathit{acc} \leftarrow \texttt{vfmacc}(\mathit{acc},\; q[i{:}] - \mathbf{base})$ \hfill // fused
  distance, no store
  \STATE \qquad $i \leftarrow i + vl$
  \STATE \quad \textbf{return} $\texttt{vfredsum}(\mathit{acc})$
}
\end{algorithmic}
\end{algorithm}

\noindent\textbf{Storage and Compute Trade-off.}
MPMI bounds per-vector storage as $\mathrm{B/v}=D + 2|\mathcal{S}_{16}| + 4|\mathcal{S}_{32}|$, directly reducing the bytes moved per candidate. Because RVV executes the fused decode path efficiently, this design trades a modest increase in vector ALU work for a much smaller memory footprint, improving the overall decode--distance balance without reintroducing the generic reshaping overhead identified in register-group-aligned widening.

\section{Reordering and Locality Optimization}
\label{sec:reorder}

To satisfy address-monotone memory access, ROrder combines global permutation with adjacency normalization at both layout levels, reshaping irregular graph probes into forward-moving address streams required by the RVV backend.


\subsection{ROrder: Traversal-Aware Index Reordering}
Modern HNSW-style search spends most stall time inside the distance kernel. On our SG2044 platform, \texttt{perf} attributes \textbf{71.08\% of sampled cache misses} to distance computation~\cite{de2010new}. This reveals a key mismatch: temporal co-visitation in graph traversal is not being converted into the spatial locality that the memory hierarchy can exploit. Graph processing on vector processors has been explored~\cite{besta2019graph}, but ANNS-specific locality optimization remains underexplored.

To mitigate this mismatch, ROrder places frequently co-accessed nodes contiguously. Generic graph orderings (e.g., RCM, BFS) do not capture ANNS-specific traversal patterns (see Sec.~\ref{sec:related} for details). We therefore use a greedy graph-reordering strategy that maximizes a locality score, aiming to produce address streams that are friendlier to caches, prefetchers, and RVV loads under irregular ANNS traversal.

Let $G=(V,E)$ denote the directed $L_0$ graph and $N^{-}(u)$ the in-neighbors of $u$.
We use two locality terms and an additive score
$S(u,v)=S_{\mathrm{s}}(u,v)+S_{\mathrm{n}}(u,v)$,
where $S_{\mathrm{s}}(u,v)=|N^{-}(u)\cap N^{-}(v)|$ counts shared in-neighbors and $S_{\mathrm{n}}(u,v)=\mathbf{1}[(u,v)\in E]+\mathbf{1}[(v,u)\in E]$ captures direct adjacency.
Here, $\mathbf{1}[\cdot]$ is the indicator function.

\noindent\textbf{Sibling Term.}
We use the sibling score $S_{\mathrm{s}}$ because HNSW traversal exhibits short-range temporal correlation from \emph{same-parent enqueuing}: if many predecessors reach both $u$ and $v$, the two are likely to be visited within a few expansions. The intersection $|N^{-}(u)\cap N^{-}(v)|$ is a topology-only proxy for this co-visitation probability. Prioritizing $S_{\mathrm{s}}$ pulls such pairs into short ID intervals, improving spatial locality and reducing the working set seen by caches and prefetchers.

\noindent\textbf{Adjacency Term.}
The adjacency score $S_{\mathrm{n}}$ captures next-hop correlation in greedy and beam search: after expanding $u$, the search often touches one of its neighbors $v$. Using $S_{\mathrm{n}}(u,v)$ places edge endpoints close in ID space, turning ''scan $u$'s neighbors $\rightarrow$ touch $v$'' into more monotone forward progress. This reduces scattered cache-line touches and creates more contiguous access opportunities.

Since locality only improves performance when accesses fall within the reach of caches, prefetchers, and downstream RVV loads, we restrict the objective to a hardware-constrained window $w$:
$F(\pi) = \sum_{i=1}^{n} \sum_{j=\max\{1,i-w\}}^{i-1} S\!\bigl(P[i],P[j]\bigr)$,
where $P[i]=\pi^{-1}(i)$, rewarding only realizable co-location.




We construct $P$ greedily from left to right, initializing $P[1]$ with a high in-degree node. Then for step $i$, we set $P[i]\!\leftarrow\!v^\star=\arg\max_{v\notin\{P[1],\ldots,P[i-1]\}} \ \sum_{j=\max\{1,i-w\}}^{i-1} S\!\bigl(v,P[j]\bigr).$


\noindent\textbf{Windowed Greedy Algorithm.}
Because global maximization is NP-hard, we use a sliding-window greedy heuristic that selects $v^\star$ at each step. At iteration $i$, the algorithm maintains a window $W_i$ of the $w$ most recently placed nodes. It selects the unplaced node $v$ that maximizes $\sum_{u\in W_i} S(v,u)$, appends $v$ to the permutation, inserts $v$ into $W_i$, and evicts the oldest element if $|W_i|{=}w$. This converts traversal co-visitation into short ID intervals for address-monotone access without modifying the edge set $E$.

\noindent\textbf{Cache-Line-Aligned Co-Layout.}
Given $\pi$, we co-locate all node-indexed arrays in a Structure-of-Arrays (SoA) layout with cache-line alignment.
For any per-node array $X[\cdot]$ (e.g., SQ and PQ codes),
$X'[\pi(u)] \leftarrow X[u]$ for all $u \in V$,
and we assign row stride $\texttt{stride}=\lceil\texttt{row\_bytes}/64\rceil\times 64$ so that each node's address is $A(t)=\texttt{base}+\texttt{stride}\cdot t$.
$A(t)$ is the starting payload address for the new ID $t$, which is strictly increasing in $t$. This monotone mapping helps reshape scattered accesses into streams more predictable for prefetchers and favorable to unit-stride or shorter-span indexed RVV loads.


\subsection{Memory Locality Optimization}
The global permutation aligns inter-node locality; however, significant inefficiency can persist within a node's adjacency. Address-monotone memory access is not fully satisfied by the global permutation alone: for each adjacency list, the payload stream seen by the RVV backend also depends on intra-list scan order. During traversal, scanning a neighbor list that is not sorted in ascending order of the permuted IDs yields a non-monotone permuted-ID sequence $t_1,t_2,\ldots,t_n$.
Under the linear address mapping $A(t)$, consecutive accesses have
$\Delta A_i = A(t_{i+1})-A(t_i)$, whose sign and magnitude vary, often resulting in jumps across cache-line and page boundaries. As a result, prefetch coverage and timeliness can degrade (stream prefetchers rely on sustained forward progress or stable strides), and cache pressure can increase due to higher replacement pressure and more frequent page crossings.

To solve this, we sort each adjacency by the new IDs produced by the global permutation. For every node $u$, $\texttt{nbrs}'(u)=\mathrm{sort}_{\uparrow}\bigl(\{\pi(v)\mid v\in N^{+}(u)\}\bigr)$. The implementation remaps neighbor IDs to $\pi(v)$ and sorts each list in ascending order. ROrder changes layout within each neighbor list, not the underlying graph, so recall remains unchanged under fixed HNSW parameters.

Under the monotone address mapping $A(t)$, ascending-ID order minimizes the sum of adjacent address deltas within each adjacency, turning neighbor scans into address-monotone streams. This improves two aspects of hardware behavior: 
(i) \emph{stream prefetching}, because forward accesses become easier to detect and prefetch;
and (ii) \emph{RVV load behavior}, because denser ID bands increase the opportunity for \texttt{vle*} loads where possible and reduce the MPKI for indexed gathers otherwise. 

In concert with global reordering, this pass improves L1/L2 utilization and provides the downstream RVV backend with denser, forward-moving access patterns that are more favorable to \texttt{LMUL}-based grouped execution and higher effective throughput for distance kernels.

\begin{algorithm}[t]
  \caption{{ROrder: Traversal-Aware Graph Reordering}}
  \label{alg:rorder}
  \small
  \begin{algorithmic}[1]
  {
  \STATE \textbf{Function} \textsc{ROrder}$(G, w, \mathit{idx})$
  \STATE \quad \textbf{Input:} HNSW $L_0$ graph $G{=}(V,E)$, window $w$, MPMI index $\mathit{idx}$
  \STATE \quad \textbf{Output:} permutation $\pi$, reordered graph $G'$
  \STATE \quad Build out-neighbor $N^{+}(u)$ and in-neighbor $N^{-}(u)$ lists from $E$
  \STATE \quad $s[v] \leftarrow |N^{-}(v)|$ for all $v$; \; init max-priority queue $H$ on $s$
  \STATE \quad $P[1] \leftarrow H.\textsc{ExtractMax}()$; \; $W \leftarrow \{P[1]\}$; 
  \STATE \quad \textsc{IncrUpdate}$(P[1], +1)$
  \STATE \quad \textbf{for} $i = 2$ \textbf{to} $|V|$ \textbf{do}
  \STATE \qquad $P[i] \leftarrow H.\textsc{ExtractMax}()$
  \STATE \qquad \textbf{if} $i > w$ \textbf{then} \; 
  \STATE \qquad \textsc{IncrUpdate}$(P[i{-}w], -1)$; \textsc{IncrUpdate}$(P[i], +1)$ 
  \STATE \qquad $W \leftarrow W \cup \{P[i]\}$
  \STATE \quad $\pi[P[i]] \leftarrow i$ for all $i$ \hfill // old-ID $\to$ new-ID
  \STATE \quad Reorder $\mathit{idx}$, graph, and $\texttt{nbrs}(u)$ by $\pi$; align to 64\,B stride
  \STATE \quad \textbf{return} $\pi, G'$
  \STATE
  \STATE \textbf{Function} \textsc{IncrUpdate}$(v, \delta)$ \hfill // incremental score update
  \STATE \quad \textbf{for} $z \in N^{+}(v)$: \; $s[z] \leftarrow s[z] + \delta$ \hfill // adjacency $S_{\mathrm{n}}$
  \STATE \quad \textbf{for} $u \in N^{-}(v)$, \textbf{for} $z \in N^{+}(u)$: \; $s[z] \leftarrow s[z] + \delta$ \hfill
  // sibling $S_{\mathrm{s}}$
  }
\end{algorithmic}
\end{algorithm}

\subsection{{ROrder Construction Analysis}}

\noindent\textbf{Time Complexity.} Here, we analyze the complexity of ROrder (Alg.~\ref{alg:rorder}). Each of the $N$ iterations calls \textsc{IncrUpdate} twice (line~11). The sibling update (line~19) costs $\sum_{u \in N^{-}(v)} |N^{+}(u)|$ per call. Summing over all nodes gives $\sum_v \sum_{u \in N^{-}(v)} |N^{+}(u)| = \sum_u d_{\text{out}}(u)^2$ (line~19), which is upper-bounded by $O(NM^2)$ under HNSW's degree bound $2M$ but is smaller in practice when graphs are sparse. Layout materialization permutes $N$ payload vectors (line~14, $O(ND)$) and sorts each of the $N$ adjacency lists (line~14, $O(NM\log M)$). The overall time complexity is $O(NM^2 + ND)$.

\textbf{Construction Acceleration.} We apply three techniques to keep the ROrder pass practical. (i)~ Incremental score updates: only the $O(M^2)$ co-visitation entries affected by the entering and leaving nodes are updated when the sliding window advances by one position. (ii)~ Priority-queue candidate selection: The greedy selection (line~9) requires extracting the highest-scoring unplaced node at each step. To accelerate that, we implement $H$ as a \emph{bucket priority queue}: all nodes are organized in a doubly-linked list sorted by integer score, with an index array \texttt{Header}$[k]$ pointing to the first and last node of score~$k$. Because each \textsc{IncrUpdate} (line~17) changes scores by ${\pm}1$, \textsc{IncrementKey} moves a node to the adjacent bucket in $O(1)$ pointer operations. Decrements use \emph{lazy counters}: a per-node \texttt{update}$[v]$ field accumulates pending $-1$'s without moving the node; settlement occurs only when the node reaches the list head during \textsc{ExtractMax} (line~9). This yields $O(1)$ amortized cost for all queue operations, avoiding the $O(\log N)$ of a binary heap. (iii)~ Parallel layout materialization:adjacency sorting (line~14) is parallelized with OpenMP across independent node neighborhoods. The resulting wall-clock build time remains practical (Table~\ref{tab:build_time}).


\section{Implementation Details}
\label{sec:impl}
\subsection{Codebase and Integration with Milvus}
\label{sec:impl-milvus}
Our prototype, {\projectname}, is implemented atop the Milvus open-source vector database by extending its computation layer (\texttt{Knowhere}) with a RISC-V Vector (RVV) backend. The released artifact contains \textbf{58,082 nonblank, non-comment source lines of code (SLOC)} after excluding third-party code.

\noindent\textbf{RVV Backend Architecture.}
Milvus \texttt{Knowhere} ships tuned SIMD backends for x86 and Arm but lacks native RVV support. We implement an RVV backend that follows existing kernel interfaces, is registered through \texttt{SIMDManager} for runtime dispatch, and uses RVV intrinsics plus the MPMI layout described in Sec.~\ref{sec:MPMI}.


\subsection{Build Configuration and Toolchain}
\label{sec:impl-build}
\textsc{\projectname} is compiled with GCC 15.1's RVV intrinsics~\cite{gcc2023}:
{
\small
\begin{lstlisting}[language=bash]
-Dknowhere_USE_RVV=ON
-march=rv64gcv_zvfhmin -mabi=lp64d -O3
\end{lstlisting}
}

The \texttt{-march=rv64gcv\_zvfhmin} flag targets RV64GC with RVV and minimal FP16 support (\texttt{ZVFHmin})~\cite{rvv2021spec}, while \texttt{-mabi=lp64d} selects the standard double-precision ABI.

\noindent\textbf{Numerical Correctness and Regression Testing.}
We validate RVV kernels against the scalar reference within tier-appropriate tolerances and check top-$k$ stability and end-to-end recall/latency regressions under identical build settings.


\subsection{Concurrency and Memory Coordination}
\label{sec:runtime}
\projectname follows Milvus's thread-per-query model, where each worker performs candidate generation and reranking. Workers are pinned to cores, hot payloads (8-bit affine bases and residual pools) are staged in cache-line-aligned SoA buffers, and a lightweight per-thread memory pool removes allocator contention at high concurrency. The ROrder pass (Sec.~\ref{sec:reorder}) places frequently co-visited vectors contiguously in ID and address space, with software prefetches at adjacency boundaries to improve timeliness. This layout reduces cache miss rates and sustains a higher SIMD utilization~\cite{malkov2018efficient}.

\vspace{-2mm}

\section{Evaluation}
\label{sec:evaluation}

\subsection{Experimental Setup}

\noindent\textbf{Hardware.}
All experiments were run on the platforms in Table~\ref{tab:hardware}.

\begin{table}[t]
\centering
\caption{Hardware platforms used in evaluation.}
\label{tab:hardware}
\scriptsize
\setlength{\tabcolsep}{2pt}
\renewcommand{\arraystretch}{1.10}

\makebox[\columnwidth][c]{%
\begin{tabularx}{\columnwidth}{@{}
  >{\hsize=1.45\hsize\raggedright\arraybackslash}X
  >{\hsize=1.05\hsize\centering\arraybackslash}X
  >{\hsize=0.55\hsize\centering\arraybackslash}X
  >{\hsize=0.55\hsize\centering\arraybackslash}X
  >{\hsize=0.75\hsize\centering\arraybackslash}X
@{}}
\hline
\textbf{Platform / System} & \textbf{CPU / SIMD} & \textbf{L1} & \textbf{L2} & \textbf{Mem} \\
\hline

\begin{tabular}[t]{@{}l@{}}
\textbf{RISC\mbox{-}V server(128b)}\\
SOPHGO SRA3-40-8
\end{tabular}
&
\begin{tabular}[t]{@{}c@{}}
64-core SG2044\\
@ 2.6\,GHz\\
\end{tabular}
&
\begin{tabular}[t]{@{}c@{}}
I/D\\
64/64KB
\end{tabular}
&
\begin{tabular}[t]{@{}c@{}}
2MB\\
/4core
\end{tabular}
&
\begin{tabular}[t]{@{}c@{}}
128GB\\
LPDDR5\\
\end{tabular}
\\
\hline

\begin{tabular}[t]{@{}l@{}}
\textbf{RISC\mbox{-}V edge(256b)}\\
Banana Pi BPI-F3
\end{tabular}
&
\begin{tabular}[t]{@{}c@{}}
8-core K1\\
@ 1.6\,GHz\\
\end{tabular}
&
\begin{tabular}[t]{@{}c@{}}
I/D\\
32/32KB
\end{tabular}
&
\begin{tabular}[t]{@{}c@{}}
512KB\\
/4core
\end{tabular}
&
\begin{tabular}[t]{@{}c@{}}
16GB\\
LPDDR4\\
\end{tabular}
\\
\hline

\begin{tabular}[t]{@{}l@{}}
\textbf{Arm server (NEON)}\\
Huawei TaiShan 2280 V2
\end{tabular}
&
\begin{tabular}[t]{@{}c@{}}
2$\times$48-core \\ 
@ 2.6\,GHz\\
\end{tabular}
&
\begin{tabular}[t]{@{}c@{}}
I/D\\
64/64KB
\end{tabular}
&
\begin{tabular}[t]{@{}c@{}}
512KB\\
/core
\end{tabular}
&
\begin{tabular}[t]{@{}c@{}}
512GB\\
DDR4\\
\end{tabular}
\\
\hline

\begin{tabular}[t]{@{}l@{}}
\textbf{Arm server (SVE)}\\
\emph{Yitian 710}
\end{tabular}
&
\begin{tabular}[t]{@{}c@{}}
64-core \\ @ 2.75\,GHz\\
\end{tabular}
&
\begin{tabular}[t]{@{}c@{}}
I/D\\
64/64KB
\end{tabular}
&
\begin{tabular}[t]{@{}c@{}}
1\,MB\\
/core
\end{tabular}
&
\begin{tabular}[t]{@{}c@{}}
128\,GB\\
DDR5\\
\end{tabular}
\\
\hline

\begin{tabular}[t]{@{}l@{}}
\textbf{x86 (AVX2 baseline)}\\
AMD EPYC 7502
\end{tabular}
&
\begin{tabular}[t]{@{}c@{}}
32-core/64thread\\
@ 2.5\,GHz
\end{tabular}
&
\begin{tabular}[t]{@{}c@{}}
I/D\\
32/32KB
\end{tabular}
&
\begin{tabular}[t]{@{}c@{}}
512KB\\
/core
\end{tabular}
&
\begin{tabular}[t]{@{}c@{}}
128GB\\
DDR4\\
\end{tabular}
\\
\hline

\begin{tabular}[t]{@{}l@{}}
\textbf{x86 (AVX\mbox{-}512 baseline)}\\
Intel Xeon 6982P\mbox{-}C
\end{tabular}
&
\begin{tabular}[t]{@{}c@{}}
64-core/64thread\\
(up to 3.9\,GHz)\\
\end{tabular}
&
\begin{tabular}[t]{@{}c@{}}
I/D\\
64/48KB
\end{tabular}
&
\begin{tabular}[t]{@{}c@{}}
2MB\\
/core
\end{tabular}
&
\begin{tabular}[t]{@{}c@{}}
128GB\\
DDR5\\
\end{tabular}
\\
\hline
\end{tabularx}%
}
\end{table}


\noindent\textbf{Characteristics of Evaluated Datasets.}
As summarized in Table~\ref{tab:dataset}, we evaluate \projectname on four public benchmarks used in Milvus: \textbf{SIFT1M} \cite{jegou2011searching}, \textbf{GIST1M} \cite{oliva2001modeling}, \textbf{BioASQ1M} \cite{tsatsaronis2015bioasq}, and \textbf{Cohere10M} \cite{cohere2023embedding}. {SIFT1M and GIST1M are widely adopted in recent architecture research~\cite{wang2021milvus}, providing representative coverage of low-dimensional (128-d) and high-dimensional (960-d) workloads. We focus detailed mechanism analysis on these two datasets while reporting end-to-end results across all four.}

\begin{table}[t]
\centering
\caption{Characteristics of the evaluated vector datasets.}
\label{tab:dataset}
\scriptsize
\begin{tabular}{lccc}
\toprule
\textbf{Dataset} & \textbf{Dimension} & \textbf{Raw Data Size} & \textbf{Data Type} \\
\midrule
SIFT1M & 128 & 493 MB & uint8  \\
\midrule
GIST1M  & 960  & 3.6 GB  & float32  \\
\midrule
BioASQ1M & 1024  & 3.9 GB & float32  \\
\midrule
Cohere10M & 768  & 44 GB & float32  \\
\bottomrule
\end{tabular}

\end{table}

\noindent\textbf{Baselines.}
We compare \projectname against five \textbf{Milvus} backends (which employ FAISS as their core search engine): 
(i) a RISC-V naive implementation (without our proposed optimizations), 
(ii) Arm SIMD backends, including NEON and SVE, and 
(iii) x86 SIMD backends, including AVX2 and AVX-512.
All baselines use identical indices and parameters to ensure a fair comparison.

\noindent\textbf{Indexes.} We evaluate HNSW and IVF across Arm, x86, and \projectname. HNSW uses $M{=}16$, $efConstruction{=}256$, and $efSearch{=}512$ unless otherwise specified; IVF uses $nlist{=}1024$ and sweeps $nprobe$ for QPS--recall trade-offs. Here, $M$ is the target number of neighbors per node; $efConstruction$ and $efSearch$ control HNSW candidate-set sizes during build and search; $nlist$ and $nprobe$ denote the number of inverted lists and probed lists in IVF. Except for Fig.~\ref{fig:qps_recall} and Fig.~\ref{fig:qps_recall_advanced}, all performance results are measured at \textbf{Recall@100 = 0.99}.

\noindent\textbf{Threading, Affinity, and NUMA Policy.} For fair comparison, we match concurrency, batch size, and memory placement across architectures. Following DiskANN~\cite{jayaram2019diskann}, we use 16 OpenMP~\cite{dagum1998openmp} threads, pin workers to physical cores, and allocate buffers locally on NUMA systems. Private-L2 differences do not dominate because thread-per-query ANNS has little inter-query sharing, and our gains reproduce on private-L2 SVE platforms (Sec.~\ref{sec:generality-isa}).

\subsection{Cross-Architecture Evaluation}
\label{sec:overall}

{We evaluate cross-architecture behavior at two scopes. First, we report the end-to-end platform-level performance against production SIMD backends; these absolute comparisons include processor, memory-hierarchy, SIMD-backend, and baseline-index differences. Second, to further isolate the proposed optimizations, we perform a controlled cross-ISA ablation that uses each platform's own SIMD+FP32 HNSW path as the reference and compares the relative gains from MPMI and ROrder.}

\begin{figure}[t]
    \centering
    \includegraphics[width=0.5\textwidth]{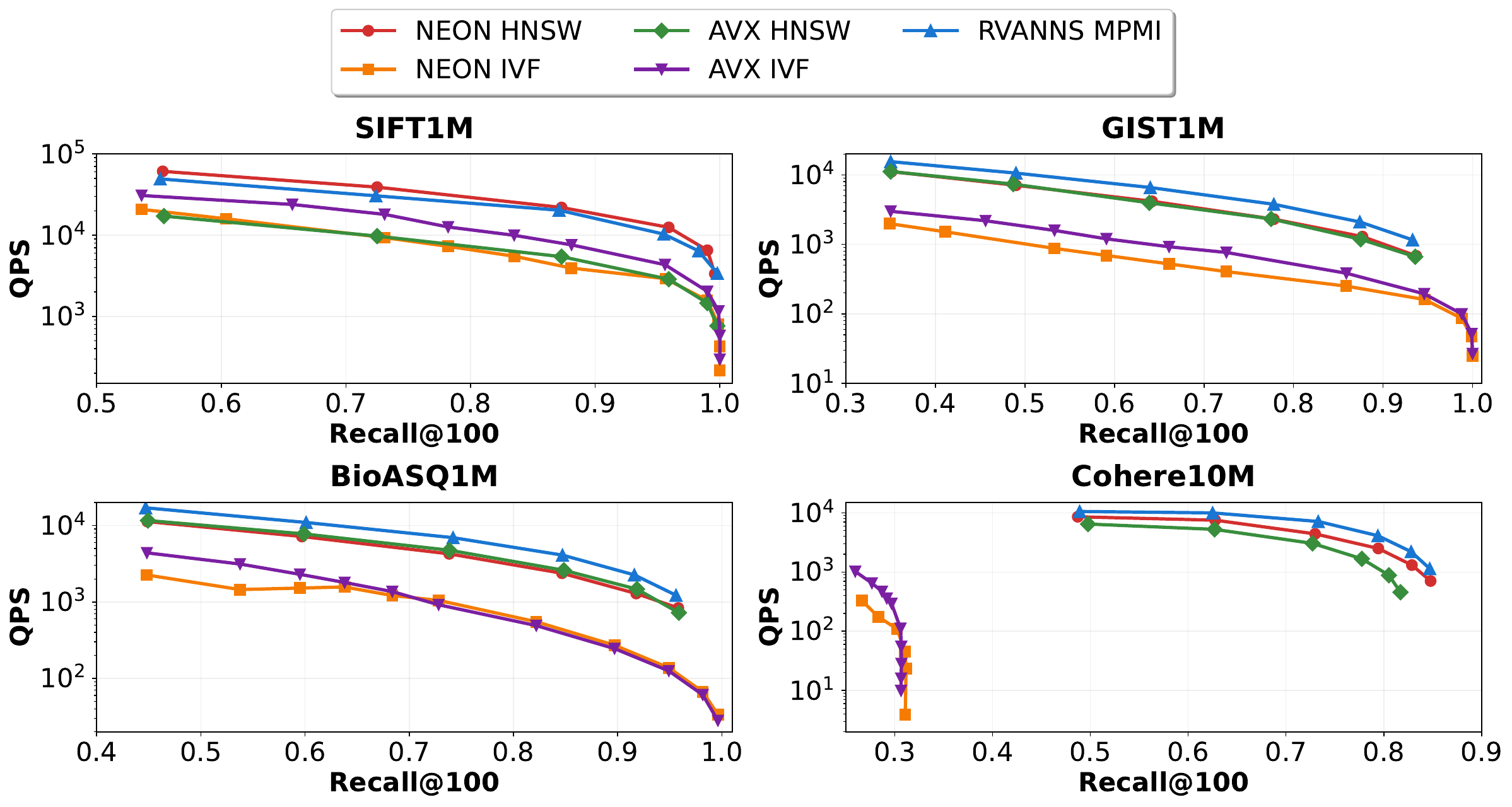}
    \caption{QPS against Arm NEON and x86 AVX2.}
    \label{fig:qps_recall}
\end{figure}

\begin{figure}[t]
    \centering
    \includegraphics[width=0.5\textwidth]{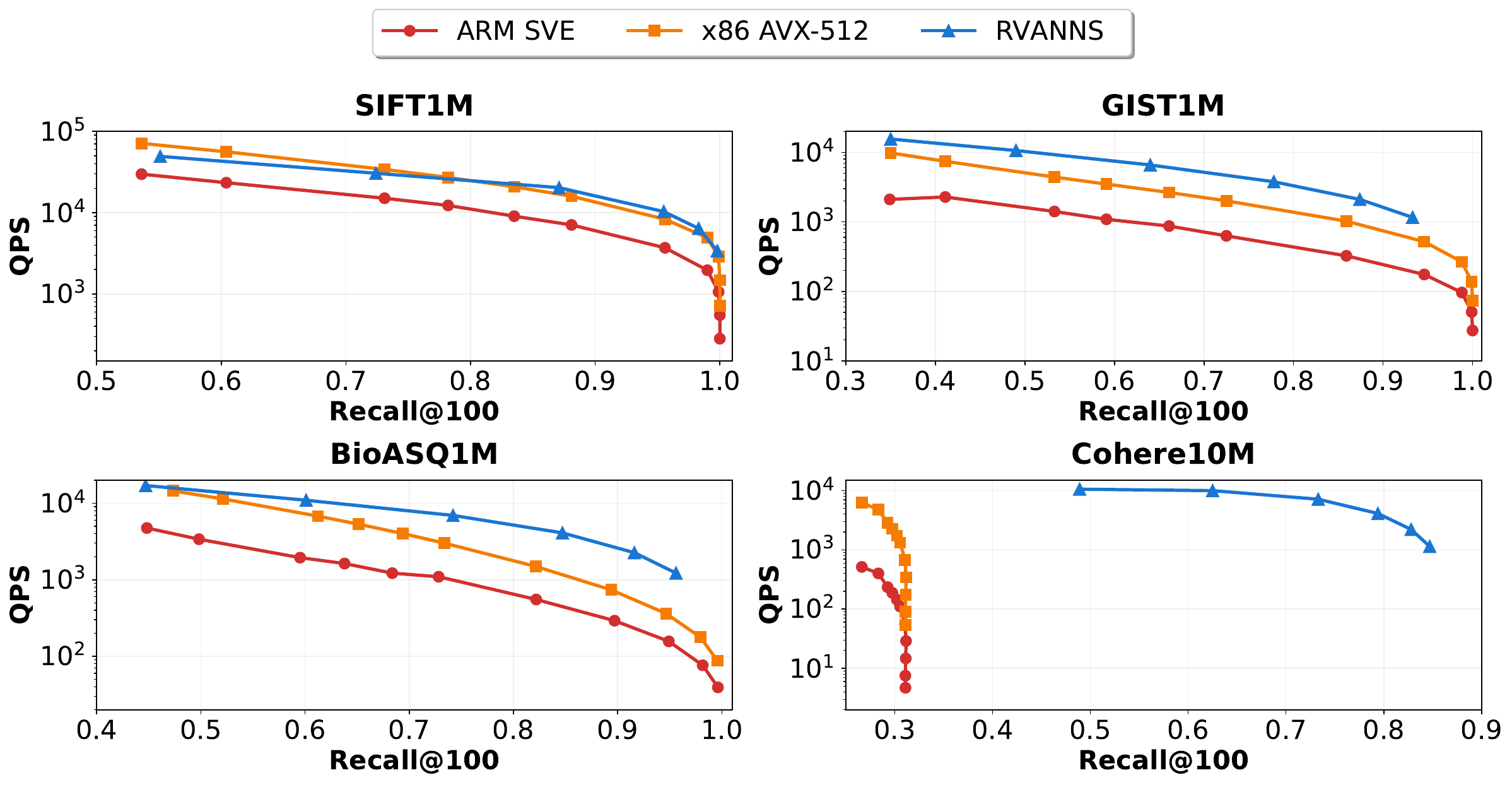}
    \caption{QPS against Arm SVE and x86 AVX-512.}
    \label{fig:qps_recall_advanced}
\end{figure}

\noindent\textbf{{End-to-End Platform Comparison.}}

{Using 16 threads and matched-recall QPS via interpolation, Fig.~\ref{fig:qps_recall} and Fig.~\ref{fig:qps_recall_advanced} compare \projectname with Arm NEON/SVE and x86 AVX2/AVX-512 on four datasets. Against fixed-width SIMD baselines, \projectname achieves up to \textbf{1.71$\times$} over NEON HNSW, \textbf{11.45$\times$} over NEON IVF, \textbf{1.39$\times$--4.44$\times$} over AVX2 HNSW, and \textbf{1.64$\times$--11.20$\times$} over AVX2 IVF. Against advanced wide-SIMD baselines, it achieves \textbf{1.74$\times$--9.31$\times$} higher QPS than SVE IVF and \textbf{1.49$\times$--3.96$\times$} higher QPS than AVX-512 IVF. These are platform-level speedups; mechanism-isolated effects of MPMI and ROrder are evaluated separately in Table~\ref{tab:generality_isa}.}

\begin{table}[t]
\centering
\caption{{Cross-ISA contribution decomposition on GIST1M and SIFT1M.
S=Scalar+FP32, V=SIMD+FP32, M=SIMD+MPMI, F=\projectname= MPMI+ROrder.
F/S, F/V, and F/M report speedups relative to scalar, SIMD FP32, and MPMI, respectively; $\Delta$ columns report V$\rightarrow$F changes.}}
\label{tab:generality_isa}
\small
\resizebox{\linewidth}{!}{%
\setlength{\tabcolsep}{2.0pt} 
\renewcommand{\arraystretch}{1.2} %
{
\resizebox{\columnwidth}{!}{%
\begin{tabular}{cc cccc ccc ccc}
\toprule
\textbf{Data} & \textbf{ISA} & \textbf{S} & \textbf{V} & \textbf{M} & \textbf{F} & \textbf{F/S} & \textbf{F/V} & \textbf{F/M} & \textbf{IPC$\Delta$} & \textbf{L1D$\Delta$} & \textbf{L2D$\Delta$} \\
\midrule
\multirow{3}{*}{GIST1M}
 & RVV     & 364.83 & 448.60  & 1005.43 & 1237.00 & \textbf{3.39$\times$} & \textbf{2.76$\times$} & \textbf{1.23$\times$} & \textbf{+70.8\%}   & \textbf{-59.1\%} & \textbf{-64.3\%} \\
 & SVE     & 719.43 & 847.48  & 1186.39 & 1351.40 & 1.88$\times$          & 1.59$\times$          & 1.14$\times$ & +61.9\% & -50.2\% & -55.5\% \\
 & AVX-512 & 865.37 & 1116.20 & 1303.14 & 1362.18 & 1.57$\times$          & 1.22$\times$          & 1.05$\times$ & +57.8\%  & -45.7\% & -46.1\% \\
\midrule
\multirow{3}{*}{SIFT1M}
 & RVV     & 1010.08 & 1460.44 & 2794.98 & 3319.76 & \textbf{3.29$\times$} & \textbf{2.27$\times$} & \textbf{1.19$\times$} & \textbf{+51.9\%} & \textbf{-63.8\%} & \textbf{-47.9\%} \\
 & SVE     & 2610.90 & 2967.28 & 3689.91 & 3861.04 & 1.48$\times$          & 1.30$\times$          & 1.05$\times$ & +32.5\% & -52.9\% & -38.5\% \\
 & AVX-512 & 2855.49 & 3324.76 & 3871.11 & 3913.56 & 1.37$\times$          & 1.18$\times$          & 1.01$\times$ & +41.2\% & -39.1\% & -28.7\% \\
\bottomrule
\end{tabular}
}
}
}
\end{table}



\noindent\textbf{{Cross-ISA Mechanism Transferability.}}
\label{sec:generality-isa}
{Table~\ref{tab:generality_isa} provides the controlled mechanism-level comparison under a common HNSW configuration on GIST1M and SIFT1M. For each ISA, V denotes the SIMD+FP32 reference; F/V measures the full MPMI+ROrder gain (F=\projectname), and F/M isolates the incremental gain from ROrder after MPMI. Across both datasets, MPMI+ROrder improves every backend: on GIST1M/SIFT1M, F/V is \textbf{2.76$\times$}/\textbf{2.27$\times$} on SG2044/RVV, \textbf{1.59$\times$}/\textbf{1.30$\times$} on Arm SVE, and \textbf{1.22$\times$}/\textbf{1.18$\times$} on x86 AVX-512. The micro-profiling columns show the same trend: from V to F, all backends improve IPC and reduce L1D/L2D MPKI, while RVV shows the strongest improvement. These results indicate that MPMI and ROrder are transferable memory-side mechanisms, but their relative gain is largest on RVV because mixed-precision decode and FP32 accumulation map to \texttt{LMUL}-grouped widening, while ROrder supplies a more locality-friendly stream to the RVV backend.}

\begin{figure}[t]
    \centering
    \includegraphics[width=0.4\textwidth]{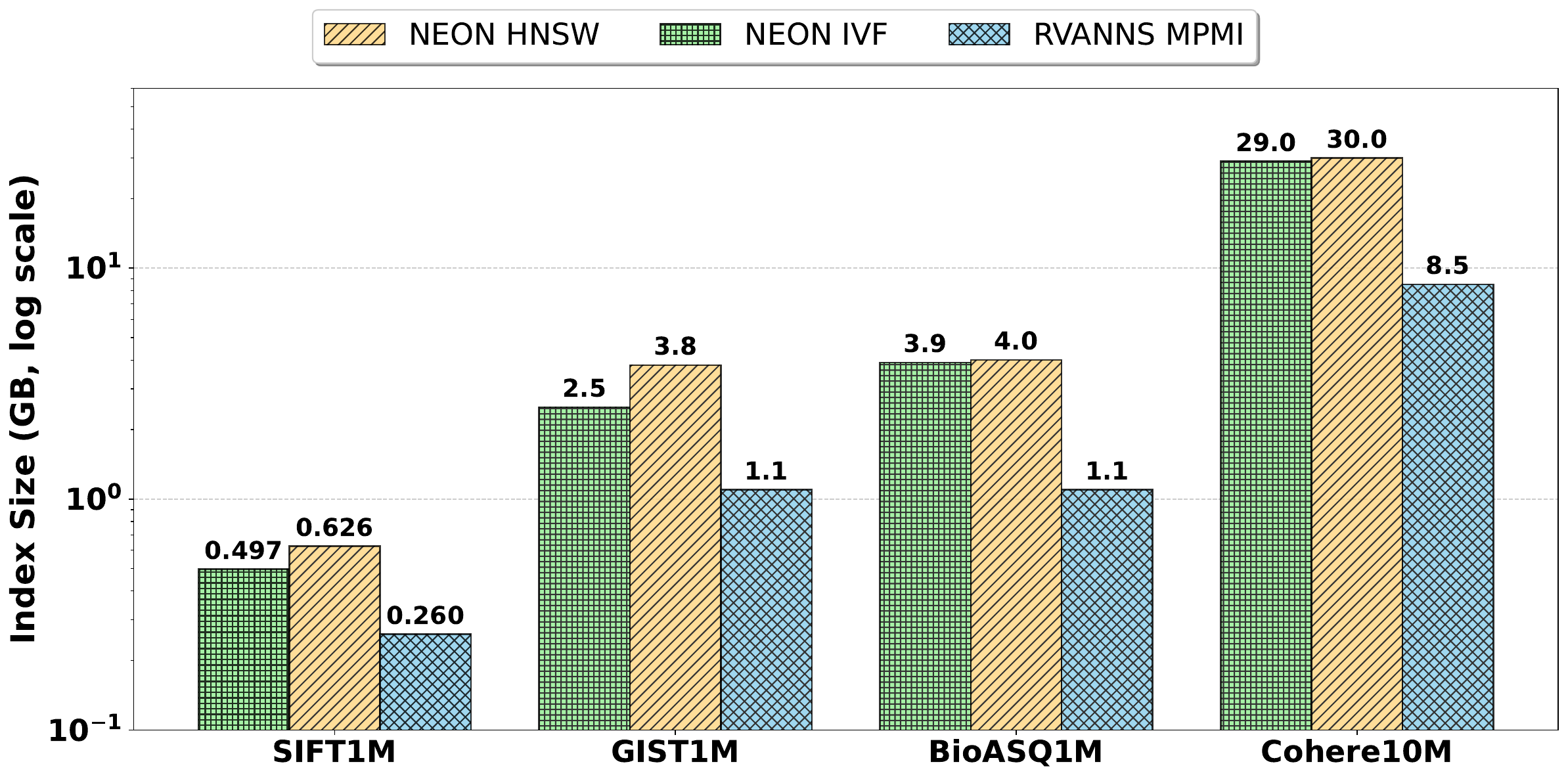}
    \caption{Index memory footprint comparison.}
    \label{fig:index_memory}
\end{figure}

\noindent\textbf{Memory Footprint.}
Fig.~\ref{fig:index_memory} reports index memory usage, where vector payloads account for up to 98\%. Baseline \texttt{HNSW} and \texttt{IVF} average 9.61\,GB and 8.97\,GB, respectively, whereas \textbf{MPMI} averages 2.74\,GB. \projectname reduces footprint by \textbf{58.5\%--72.5\%} over NEON \texttt{HNSW} and \textbf{47.7\%--71.8\%} over NEON \texttt{IVF}, validating MPMI's ability to reduce bytes moved per candidate and ease bandwidth pressure during distance evaluation.


\subsection{Thread Scalability}
\label{sec:thread_scaling}

\begin{figure}[t]
    \centering
    \includegraphics[width=\columnwidth]{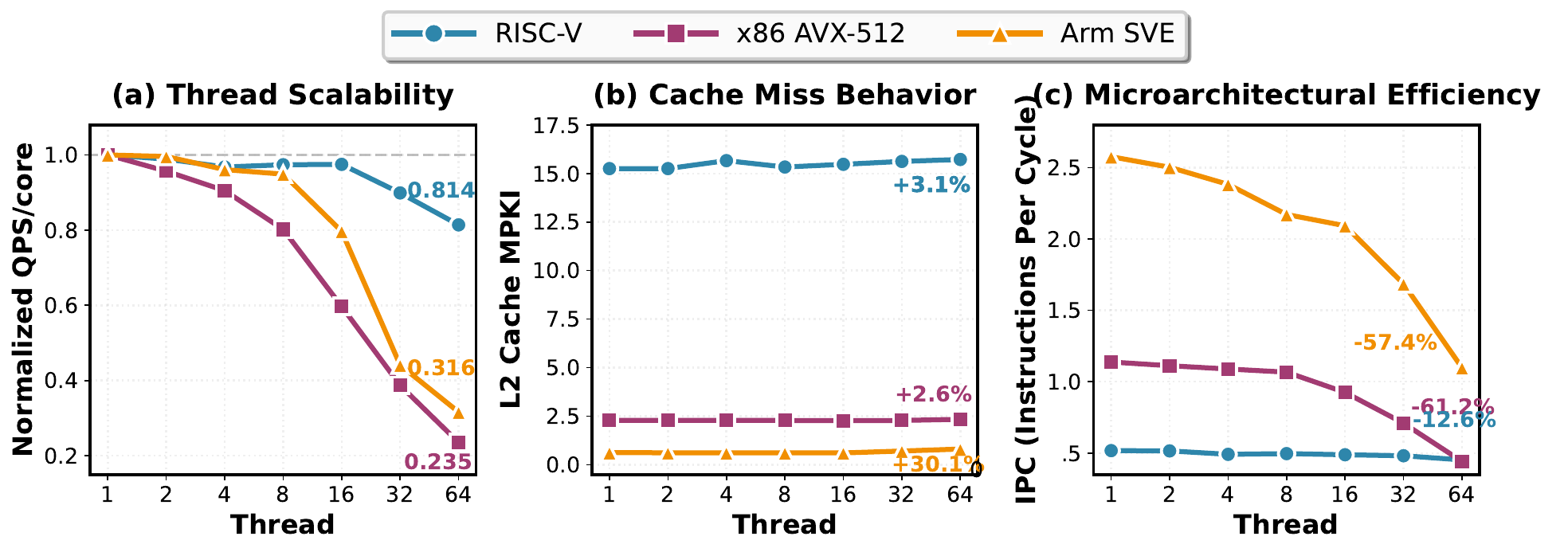}
    \caption{{
    Cross-ISA thread scalability on GIST1M.
    (a) Normalized per-core throughput, $\frac{Q(T)}{T\cdot Q(1)}$;
    (b) Cache miss, L2D MPKI.
    (c) Microarchitectural efficiency, IPC.
    }}
    \label{fig:thread_scaling_bottleneck}
\end{figure}

{
Fig.~\ref{fig:thread_scaling_bottleneck} analyzes thread scalability from two levels: normalized per-core throughput $\frac{Q(T)}{T \cdot Q(1)}$, and micro-profiling indicators including L2D MPKI and IPC. At 64 threads, RVV achieves 0.814 normalized QPS/Core, outperforming AVX-512 (0.235, 3.5$\times$ lower) and SVE (0.316, 2.6$\times$ lower). This high-level trend is consistent with Table~\ref{tab:generality_isa}.

The micro-profiling views explain this scalability gap. With MPMI+ROrder, RVV maintains stable cache behavior under concurrency: its L2D MPKI grows only 3.1\% (15.25 $\rightarrow$ 15.72), showing that the reordered payload stream preserves locality as thread count increases. At the same time, RVV retains higher execution efficiency: IPC drops by only 12.6\% (0.517 $\rightarrow$ 0.452), whereas AVX-512 drops by 61.2\% (1.138 $\rightarrow$ 0.441) and SVE by 57.4\% (2.577 $\rightarrow$ 1.098).

The cross-ISA contrast shows that stable cache behavior is insufficient. AVX-512 has the most stable L2D MPKI (+2.6\%) but the worst QPS/Core scaling, indicating execution-side saturation rather than a bandwidth limit. SVE suffers both cache degradation (+30.1\% MPKI) and execution inefficiency (-57.4\% IPC). In contrast, RVV combines stable memory behavior with low execution overhead: its \texttt{vsetvl}-based VLA execution and \texttt{LMUL}-grouped register allocation better convert the locality-friendly stream from MPMI+ROrder into scalable per-core throughput under high thread counts.
}

\subsection{VLA Execution Efficiency Across VLENs}
\label{sec:generality-vlen}

Although VLA execution is also used by Arm SVE, this subsection isolates the RVV execution substrate used by our backend. We evaluate both the efficiency of RVV's \texttt{vsetvl}-based execution and its portability across different \texttt{VLEN} configurations.

\begin{table}[t]
\centering
\caption{Throughput ablation of VLA execution.} 
\label{tab:vla_threeway}
\small
\begin{tabular}{lccc}
\hline
\textbf{Configuration} & \textbf{QPS} & \textbf{Speedup} & \textbf{vs.\ Prev.} \\
\hline
Scalar & 364.83 & 1.00$\times$ & -- \\
Fixed-Length & 383.07 & \textbf{1.05$\times$} & \textbf{+5.00\%}  \\
VLA SIMD & 448.60 & \textbf{1.23$\times$} & \textbf{+17.11\%} \\
\hline
\end{tabular}
\end{table}

\noindent\textbf{VLA Ablation on Efficiency.}
To isolate the contribution of RVV's VLA execution, we compare three configurations on GIST1M ($D{=}960$) with fixed HNSW parameters: (1)~\textbf{Scalar}, (2)~\textbf{Fixed-Length} (\texttt{vl}{=}4 FP32 elements with software-managed strip-mining and control handling), and (3)~\textbf{VLA SIMD} (\texttt{vsetvl}-based). As shown in Table~\ref{tab:vla_threeway}, Fixed-Length achieves \textbf{1.05$\times$} speedup over Scalar, while VLA SIMD reaches \textbf{1.23$\times$}. Even though $D{=}960$ is a vector width multiple, VLA still improves performance by simplifying the control path, reducing strip-mining overhead, and improving downstream scheduling.

\noindent{\textbf{Scalability Across VLENs.}
We evaluate two real RVV implementations with different vector lengths: SG2044 (\texttt{VLEN}{=}128\,b) and Banana Pi BPI-F3 (\texttt{VLEN}{=}256\,b). The same VLA RVV binary runs on both systems without recompilation, exercising RVV's vector-length-agnostic execution while keeping the software artifact fixed. Table~\ref{tab:vlen_generality} reports within-platform speedups on GIST1M. SIMD+FP32 improves over Scalar by \textbf{1.23$\times$} on SG2044 and \textbf{1.36$\times$} on BPI-F3, showing that VLA vector execution provides portable baseline acceleration. Adding MPMI+ROrder on top of SIMD+FP32 further improves throughput by \textbf{2.76$\times$} and \textbf{3.63$\times$}, yielding end-to-end speedups of \textbf{3.39$\times$} and \textbf{4.94$\times$} over Scalar. These results show that the same VLA binary remains effective across the measured 128\,b and 256\,b RVV configurations. Since the two systems differ in cores, caches, and memory systems, we interpret this table through within-platform ratios rather than cross-platform absolute QPS.}

\begin{table}[t]
\centering
\caption{{Generality across measured RVV VLENs on GIST1M. Speedups are ratios: V/S = SIMD+FP32/Scalar, F/V = \projectname/SIMD+FP32, and F/S = \projectname/Scalar.}}
\label{tab:vlen_generality}
{
\small
\resizebox{\linewidth}{!}{
\begin{tabular}{lccccccc}
\toprule
\textbf{Platform} & \textbf{VLEN} & \textbf{Scalar} & \textbf{SIMD+FP32} & \textbf{\projectname} & \textbf{V/S} & \textbf{F/V} & \textbf{F/S} \\
\midrule
SG2044 & 128\,b & 364.83 & 448.60 & 1237.00 & 1.23$\times$ & 2.76$\times$ & \textbf{3.39$\times$} \\
BPI-F3 & 256\,b & 34.64 & 47.15 & 171.16 & 1.36$\times$ & 3.63$\times$ & \textbf{4.94$\times$} \\
\bottomrule
\end{tabular}
}
}
\end{table}

\begin{table}[t]
\centering
\caption{Decode-path microbenchmark for Challenge~1.}
\label{tab:decode_micro}
\small
\begin{tabular}{llcc}
\toprule
\textbf{\texttt{LMUL}} & \textbf{Decode Path} & \textbf{Cycles (B)} & \textbf{Instr. (B)} \\
\midrule
\texttt{LMUL=1} & Conventional SQ8 & 52.88 & 51.22 \\
\texttt{LMUL=1} & MPMI Fused & 49.50 \textbf{(-6.4\%)} & 31.04 \textbf{(-39.4\%)} \\
\midrule
\texttt{LMUL=4} & Conventional SQ8 & 32.59 & 19.12 \\
\texttt{LMUL=4} & MPMI Fused & \textbf{23.11} \textbf{(-29.1\%)} & \textbf{9.36} \textbf{(-51.1\%)} \\
\bottomrule
\end{tabular}
\end{table}

\begin{table}[t]
\centering
\small
\caption{{Full-path RVV \texttt{LMUL} sensitivity on GIST1M. M=MPMI and F=MPMI+ROrder; $\Delta$ report M$\rightarrow$F changes.}}
\label{tab:rvv_grouping_full_path}

{
\resizebox{\linewidth}{!}{
\begin{tabular}{lcccccc}
\toprule
\textbf{\texttt{LMUL}} & \textbf{M QPS} & \textbf{F QPS} & \textbf{F/M} & \textbf{Cycles$\Delta$} & \textbf{IPC$\Delta$} & \textbf{L2D MPKI$\Delta$} \\
\midrule
\texttt{no-LMUL} & 55.97 & 57.30 & 1.02$\times$ & -3.4\% & +3.9\% & -3.3\% \\
\texttt{LMUL=1} & 59.07 & 61.58 & 1.04$\times$ & -4.4\% & +4.5\% & -6.6\% \\
\texttt{LMUL=4} & 60.31 & 63.97 & \textbf{1.06$\times$} & \textbf{-6.2\%} & \textbf{+6.5\%} & \textbf{-9.3\%} \\
\texttt{LMUL=8} & 58.91 & 61.51 & 1.04$\times$ & -5.3\% & +4.4\% & -6.6\% \\
\bottomrule
\end{tabular}
}
}
\end{table}

\begin{table}[t]
\centering
\caption{System-level impact of MPMI compared with HNSW and IVF within \projectname.}
\label{mpmi_impact}
\small
\resizebox{\linewidth}{!}{
\begin{tabular}{lcccccc}
\toprule
\textbf{Metric} & \textbf{Dataset} & \textbf{HNSW} & \textbf{IVF} & \textbf{MPMI} & \textbf{vs. HNSW} & \textbf{vs. IVF} \\
\midrule
{\textbf{QPS}}  
& SIFT1M & 1460.44 & 846.52 & 2794.98 & \textbf{+91.38\%} $\uparrow$ & \textbf{+230.14\%} $\uparrow$ \\
& GIST1M & 448.60 & 118.12 & 1005.43 & \textbf{+124.14\%} $\uparrow$ & \textbf{+751.02\%} $\uparrow$ \\
\midrule
\textbf{L1D MPKI}
& SIFT1M & 8.53 & 15.19 & 3.85 & \textbf{-54.87\%} $\downarrow$ & \textbf{-74.66\%} $\downarrow$ \\
& GIST1M & 9.22 & 10.47 & 4.33 & \textbf{-53.04\%} $\downarrow$ & \textbf{-58.64\%} $\downarrow$ \\
\midrule
\textbf{L2D MPKI}
& SIFT1M & 40.55 & 35.59 & 26.55 & \textbf{-34.52\%} $\downarrow$ & \textbf{-25.40\%} $\downarrow$ \\
& GIST1M & 29.70 & 27.00 & 11.23 & \textbf{-62.19\%} $\downarrow$ & \textbf{-58.41\%} $\downarrow$ \\
\bottomrule
\end{tabular}
}
\end{table}

\subsection{Efficiency of MPMI}
\label{sec:eval_mpmi}
MPMI reduces decode-side data movement while preserving RVV's register-group-aligned widening. We evaluate it through kernel-level decode profiling, full-path \texttt{LMUL} sensitivity, and end-to-end QPS/cache behavior.

\noindent\textbf{RVV Co-Design Effect.} Table~\ref{tab:decode_micro} compares Conventional SQ8 and MPMI Fused under matched decode-and-distance settings. Conventional SQ8 materializes an FP32 buffer before distance evaluation, whereas MPMI Fused streams 8-bit base reconstruction, residual fusion, and accumulation in the same \texttt{vfmacc} loop. MPMI Fused consistently reduces decode overhead under the same \texttt{LMUL}: at \texttt{LMUL=1}, it reduces cycles by \textbf{6.4\%} and retired instructions by \textbf{39.4\%}; at \texttt{LMUL=4}, it reduces cycles by \textbf{29.1\%} and retired instructions by \textbf{51.1\%}. These gains show that MPMI aligns mixed-precision decode with \texttt{LMUL}-based widening while avoiding FP32 materialization.

\noindent\textbf{{Full-Path RVV \texttt{LMUL} Sensitivity.}} 
{ Table~\ref{tab:rvv_grouping_full_path} evaluates the full MPMI search path under different \texttt{LMUL} settings on GIST1M with one thread at matched recall. The \texttt{no-LMUL} row disables explicit RVV grouped decode and accumulation while keeping the same MPMI representation and layout. Explicit \texttt{LMUL} improves both MPMI and MPMI+ROrder, with \texttt{LMUL=4} achieving the highest QPS; \texttt{LMUL=8} remains faster than \texttt{no-LMUL} but below \texttt{LMUL=4}, reflecting the trade-off between larger \texttt{VLMAX} and higher register pressure. ROrder improves every \texttt{LMUL} setting on top of MPMI, increasing QPS by 1.02--1.06$\times$, reducing cycles by 3.4--6.2\%, improving IPC by 3.9--6.5\%, and lowering L2D MPKI by 3.3--9.3\%. This confirms that MPMI preserves grouped RVV execution benefits, while ROrder further reduces traversal-side cache-miss pressure. }

\noindent\textbf{Comparison with HNSW and IVF.} Table~\ref{mpmi_impact} evaluates the system-level impact of MPMI within the same \projectname framework. On SIFT1M, MPMI delivers \textbf{1.35$\times$} higher QPS than HNSW and \textbf{3.43$\times$} higher QPS than IVF; on GIST1M, it improves QPS by \textbf{1.20$\times$} over HNSW and \textbf{9.12$\times$} over IVF. These gains show that reducing bytes moved per candidate does not shift the bottleneck into decode. MPMI reduces the batch-4 kernel working set from 15.36\,KB to 3.84\,KB, making the decode path more cache-resident and reducing cache miss intensity, including up to \textbf{54.87\%} lower L1D MPKI on SIFT1M and \textbf{62.19\%} lower L2D MPKI on GIST1M relative to HNSW.

\subsection{Efficiency of ROrder}
\label{ablation_3}

ROrder improves search-time locality by co-locating likely co-visited nodes and sorting adjacency lists by reordered IDs. We evaluate its direct impact on the RVV+MPMI search pipeline and compare it with graph-reordering baselines.

\noindent\textbf{RVV Co-Design Effect.}
Table~\ref{tab:generality_isa} evaluates ROrder on top of the MPMI pipeline at Recall@100{=}0.99 on SIFT1M. As shown in Sec.~\ref{sec:motivation}, \texttt{LMUL}-grouped RVV execution covers more candidate work per vector instruction, but irregular graph traversal still fetches candidates from scattered payload addresses. ROrder targets this bottleneck from the layout side: it reshapes the address stream feeding the decode-and-distance kernel by co-locating co-visited nodes and sorting adjacency lists by reordered IDs.
Applying ROrder increases throughput by \textbf{18.77\%} (QPS: 2794.98 $\rightarrow$ 3319.76), while reducing L1D MPKI by \textbf{19.97\%} (3.85 $\rightarrow$ 3.08) and L2D MPKI by \textbf{20.41\%} (26.55 $\rightarrow$ 21.13). It also reduces per-query memory traffic by \textbf{14.2\%} and aggregate bandwidth demand from 28.48 to 27.20\,GB/s. These results show that ROrder reduces traversal-side cache-miss pressure and translates layout locality into higher end-to-end QPS.







\begin{table}[t]
\centering
\caption{Comparison of graph orderings.}
\label{tab:rorder_ordering_compare}
\scriptsize
{
\setlength{\tabcolsep}{2pt}
\renewcommand{\arraystretch}{0.95}
\resizebox{\columnwidth}{!}{
\begin{tabular}{lrrrrrr}
\hline
\textbf{Metric} & \textbf{None} & \textbf{BFS} & \textbf{GOrder} & \textbf{RCM} & \textbf{RabbitOrder} & \textbf{ROrder} \\
\hline
QPS          & 2794.98 & 2891.03 & 3018.73 & 2990.94 & 2866.58 & \textbf{3319.76} \\
L1D MPKI     & 3.85    & 3.52    & 3.34    & 4.14    & 4.60    & \textbf{3.08} \\
L2D MPKI     & 26.55   & 23.22   & 21.74   & 22.96   & 23.86   & \textbf{21.13} \\
\hline
\end{tabular}
}
}
\end{table}

\noindent\textbf{Comparison with Graph-Reordering Baselines.}
Table~\ref{tab:rorder_ordering_compare} compares ROrder with representative graph-reordering methods under the same graph, recall target, and MPMI pipeline. Generic orderings improve structural graph locality, but they do not directly optimize the priority-ordered candidate stream produced by ANN beam search. BFS, RCM~\cite{cuthill1969reducing}, and RabbitOrder~\cite{arai2016rabbitorder} yield modest QPS gains of 3.4\%, 7.0\%, and 2.6\%, respectively. GOrder is the strongest generic baseline, increasing QPS to 3018.73 and reducing L1D/L2D MPKI to 3.34/21.74. ROrder further aligns layout with query-time co-visitation through its sibling-adjacency score and reordered adjacency lists, achieving the highest QPS (\textbf{3319.76}) and lowest L1D/L2D MPKI (\textbf{3.08}/\textbf{21.13}). Overall, ROrder improves throughput by \textbf{18.77\%} over the unordered layout and by \textbf{9.97\%} over GOrder, showing that traversal-aware layout better matches ANN search than generic graph ordering.

\begin{table}[t]
\centering
\caption{Execution time breakdown.}
\label{tab:execution_time}
\small
\begin{tabular}{lcc}
\hline
\textbf{Component} & \textbf{SIFT1M (128-dim)} & \textbf{GIST1M (960-dim)} \\
\hline
Decode & 15.60\% & 38.00\% \\
Distance Calc & 37.20\% & 43.50\% \\
Graph Traversal & 47.20\% & 18.50\% \\
\hline
\end{tabular}
\end{table}

\subsection{Compute-Memory Trade-off Analysis}
To understand how co-design redistributes execution costs, Table~\ref{tab:execution_time} reports the latency breakdown of the MPMI-based query path. On SIFT1M, graph traversal and decode\,+\,distance contribute 47.20\% and 52.80\% of total time, respectively; on GIST1M, decode-and-distance evaluation accounts for \textbf{81.50\%} of execution cycles.


This shift does not imply that the high-dimensional workload becomes compute-bound. Consistent with the SG2044 PMU profile in Fig.~\ref{fig:anns_time_distribution}, the decode\,+\,distance path remains limited by memory-side service: the counter trend is high L2D MPKI with low IPC, not a high branch-miss rate or instruction-supply-only stall. The 38.00\% decode share on GIST1M instead reflects a compute--memory trade-off: 8-bit compression reduces the base payload by \textbf{4$\times$}, and the added reconstruction arithmetic is amortized by fewer cache misses and lower bandwidth pressure. Because vectorized decoding on RVV is much cheaper than a DRAM round trip, this trade-off lowers end-to-end latency rather than creating a new compute bottleneck.


\subsection{Parameter Sensitivity}
\label{sec:param_sensitivity}

\begin{figure}[t]
    \centering
    \includegraphics[width=\linewidth]{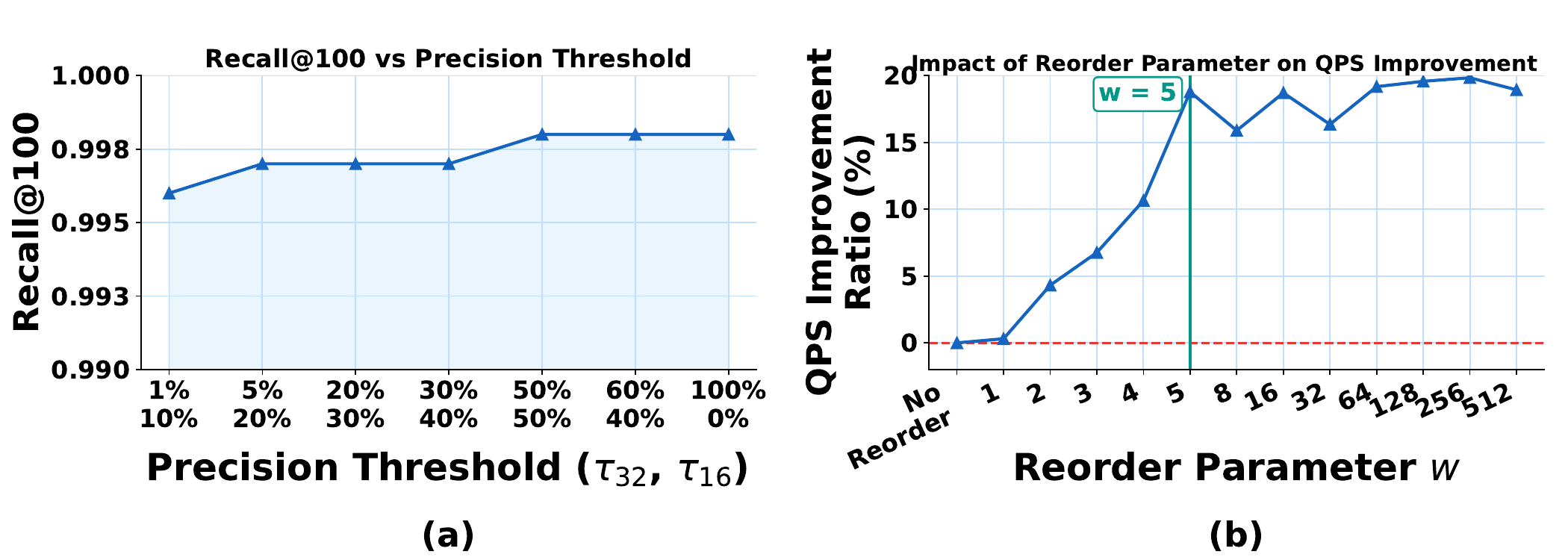}
    \caption{Parameter sensitivity experiments.}
    \label{fig:sensitivity}
\end{figure}

Fig.~\ref{fig:sensitivity} shows robust, near-optimal performance across broad ranges of ROrder window $w$ and MPMI thresholds $(\tau_{32}, \tau_{16})$.

\noindent\textbf{ROrder Window Size Sensitivity.}
ROrder uses a windowed greedy procedure; larger $w$ expands the search 
space and increases offline cost roughly linearly. Fig.~\ref{fig:sensitivity}(b) reports results on SIFT1M dataset. QPS increases from 2794.98 (no reordering) to 3319.76 at $w{=}5$ (+18.77\%), and $w{=}5$ reaches 99.2\% of the best observed QPS across all tested values. This trend is consistent with ROrder's objective: local graph structure dominates $S(u,v)$, so enlarging the window beyond a moderate size yields diminishing returns. We therefore use $w{=}5$ by default.

\noindent\textbf{MPMI Residual Thresholds Sensitivity.}
Fig.~\ref{fig:sensitivity}(a) evaluates sensitivity to $(\tau_{32},\tau_{16})$ on SIFT1M, ranging from $(1\%,10\%)$ to $(100\%,0\%)$. Across this sweep, Recall@100 remain between 0.996 and 0.998, indicating that the 8-bit affine base already preserves most dimensions with sufficient fidelity. We therefore adopt $(\tau_{32},\tau_{16}){=}(1\%,10\%)$ by default, which provides the highest compression ratio with negligible recall loss ($<$0.002) relative to the fully FP32 baseline.

\subsection{Offline Cost: Index Build Time}

\begin{table}[t]
\centering
\caption{{Comparison of index build time across datasets.}}
\label{tab:build_time}
\footnotesize
\resizebox{\columnwidth}{!}{%
\setlength{\tabcolsep}{3pt}
\renewcommand{\arraystretch}{1.10}
{
\begin{tabular}{@{}l rrr r rr cc@{}}
\toprule
\textbf{Dataset} & \textbf{MPMI} & \textbf{ROrder} & \textbf{Total} & \makecell{\textbf{ROrder/}\\\textbf{Total}} & \textbf{HNSW} & \textbf{IVF} & \makecell{\textbf{HNSW/}\\\textbf{MPMI}} & \makecell{\textbf{HNSW/}\\\textbf{Total}} \\
\midrule
SIFT1M    &   56.8s &  54.6s &  111.4s & 49.0\% &   80.2s &  82.1s & 1.41$\times$ & 0.72$\times$ \\
GIST1M    &  154.5s &  36.3s &  190.8s & 19.0\% &  275.3s & 590.2s & 1.78$\times$ & 1.44$\times$ \\
BioASQ1M  &  151.9s &  61.7s &  213.6s & 28.9\% &  299.1s & 161.9s & 1.97$\times$ & 1.40$\times$ \\
Cohere10M & 1423.7s & 689.0s & 2112.7s & 32.6\% & 2709.3s & 443.8s & 1.90$\times$ & 1.28$\times$ \\
\bottomrule
\end{tabular}
}
}%
\end{table}

Table~\ref{tab:build_time} reports the offline preprocessing overhead.

\noindent\textbf{MPMI Graph Construction.}
MPMI achieves \textbf{1.41--1.97$\times$} build speedup over HNSW because its compact codes reduce per-vector footprint during graph construction, improving cache residency of candidate vectors. IVF builds faster on BioASQ1M and Cohere10M because it avoids graph construction, but its lower query throughput at matched recall means faster indexing does not necessarily translate into better end-to-end efficiency.

\noindent\textbf{ROrder Preprocessing Cost.}
ROrder accounts for \textbf{19\%--49\%} of total indexing time. As analyzed in Sec.~\ref{sec:reorder}, its permutation phase runs in $O(\sum_u d(u)^2)$ time, upper-bounded by $O(NM^2+ND)$, and depends on graph topology rather than dimensionality $D$. Thus, GIST1M can take less ROrder time than SIFT1M despite its higher dimension. The one-time cost is amortized by query-time gains; on Cohere10M, the 689\,s preprocessing cost recover after ${\sim}10^7$ queries.

\noindent\textbf{End-to-End Comparison (Build + Query).}
Accounting for index construction plus \(10^{7}\) queries at matched high recall, \projectname{} is on par with HNSW on SIFT1M and \textbf{1.52$\times$--1.71$\times$} faster on the other three datasets, due to both faster indexing and higher query throughput. Although IVF often builds faster, \projectname{} still achieves \textbf{3.07$\times$--9.73$\times$} better end-to-end performance on SIFT1M, GIST1M, and BioASQ1M, where IVF attains comparable recall. IVF does not reach comparable recall on Cohere10M (3.43 QPS at Recall@100 = 0.32) and is omitted from this comparison.

\subsection{{Comparison with GPU Baselines}}
\label{sec:gpu_power_compare}

{
We add a CPU/GPU comparison on Cohere10M using one NVIDIA RTX A6000 and one NVIDIA A100 GPU. GPU baselines use FAISS-GPU~\cite{johnson2017faiss} and NVIDIA cuVS~\cite{nvidia2026cuvs}, with $k{=}100$, batch size 1000, and repeat 100. \projectname runs the SIMD+MPMI+ROrder path on SG2044. We measure SG2044 platform power over PSU1+PSU2 using \texttt{ipmitool}, following recent SG2044 characterization~\cite{venieri2026montecimonev3}; GPU baselines report NVIDIA device power. }

\begin{table}[t]
\centering
\caption{{Comparison of \projectname{} with GPU baselines on Cohere10M regarding QPS, device power, energy per query ($J/query$), and energy efficiency ($QPS/W$).}}
\label{tab:gpu_power_compare}
\small
\setlength{\tabcolsep}{4pt}
\renewcommand{\arraystretch}{1.08}
{
\begin{tabular}{@{}llcccc@{}}
\toprule
\textbf{Platform} & \textbf{Backend} & \textbf{QPS} & \textbf{W} & \textbf{J/q} & \textbf{QPS/W} \\
\midrule
SG2044 & \projectname & 620.48 & 84.71 & \textbf{0.1365} & \textbf{7.32} \\
A6000 & cuVS & 957.57 & 297.86 & 0.3110 & 3.22 \\
A6000 & FAISS-GPU & 1084.37 & 299.14 & 0.2759 & 3.62 \\
A100 & cuVS & 925.54 & 242.73 & 0.2623 & 3.81 \\
A100 & FAISS-GPU & 991.99 & 246.87 & 0.2489 & 4.02\\
\bottomrule
\end{tabular}
}
\end{table}

{
At matched recall, GPU baselines provide higher throughput, reaching 925.54--1084.37 QPS, while \projectname reaches 620.48 QPS. However, \projectname achieves better energy efficiency under the device-power scope: 7.32 QPS/W and 0.1365 J/query, corresponding to \textbf{1.82--2.27$\times$} higher QPS/W and \textbf{45.2--56.1\%} lower energy per query than the GPU rows. These results identify a complementary operating point. GPUs remain preferable for maximum batched throughput and device-resident search, but they are less power-efficient in this setting. In contrast, \projectname is attractive for CPU-resident vector databases that benefit from high host-memory capacity, simpler software integration, and lower power cost.
}

\subsection{Discussion: Generality and Limitations}
\label{sec:general_discussion}
The evaluation supports three main conclusions. First, RVV's VLA execution provides a portable software substrate across \texttt{VLEN} configurations, but vectorization alone does not dissolve the memory wall. Second, the dominant end-to-end gains stem from our co-designed mechanisms: MPMI reduces memory traffic and decode working sets, while ROrder improves traversal locality to minimize miss-induced stalls. Third, while these benefits partially transfer to legacy SIMD backends, the strongest gains manifest when paired with the intended RVV execution model.

At the same time, the cross-ISA results should be interpreted as platform-level transferability evidence rather than a cycle-for-cycle isolation of ISA effects, since the evaluated processors also differ in frequency, cache hierarchy, and memory system. In addition, the VLA efficiency study uses a width-aligned dataset; evaluating non-aligned dimensions would characterize tail-handling effects more explicitly. Finally, because ROrder is an offline graph transformation, workloads with frequent dynamic insertions or streaming updates remain an important direction for future work.


\section{Related Work}
\label{sec:related}

\noindent\textbf{SIMD-Accelerated ANNS Systems.}
Modern vector databases such as \textbf{FAISS}~\cite{johnson2017faiss}, \textbf{Milvus}~\cite{wang2021milvus}, and \textbf{ScaNN}~\cite{guo2020accelerating} rely on fixed-width SIMD (AVX2/NEON) or predicate-centric SVE, incurring tail-handling or predicate-management overhead on irregular ANNS workloads. Prior sparse and approximate SIMD works target fixed hardware, limiting portability. By contrast, \projectname addresses these limitations through hardware-software co-design.

\noindent\textbf{Memory Locality Optimization for ANNS.}
Prior works mitigate irregular ANNS access via locality optimization. Reordering to improve locality is also common in graph traversal and sparse computations; ROrder shares this goal, but it is specialized for ANNS where query time is dominated by vector-payload accesses in distance computation. GOrder~\cite{wei2016gorder} uses generic graph metrics that do not explicitly model ANNS-specific traversal patterns. RCM \cite{cuthill1969reducing} reduces matrix bandwidth through BFS-based reordering, while RabbitOrder \cite{arai2016rabbitorder} exploits hierarchical community structure. In contrast, neither method targets query-time co-visitation patterns in ANN graph traversal. DiskANN~\cite{jayaram2019diskann} proposes disk-aware graph layouts. Our ROrder uses a sibling-adjacency locality score with a windowed greedy construction to cluster co-visited nodes, and materializes them in a 64-byte-aligned structure-of-arrays to improve forward-biased address streams.

\noindent\textbf{RISC-V Vector Applications and Generality.}
While RVV is gaining traction in HPC~\cite{platzer2021ten}, deep learning~\cite{gonzalez2022openvla}, and compilers~\cite{ye2021scalehls,chen2018tvm}, data-intensive ANNS remains underexplored~\cite{xu2022xiangshan}. {\projectname} is the first to systematically co-design ANN indexing with RVV's semantics. ROrder's locality transformation is broadly portable, while MPMI's memory-side benefits partially transfer to advanced SIMD backends with efficient gather and predication support (e.g., evaluated on AVX-512 and SVE). Crucially, both mechanisms are nevertheless RVV-oriented: their design is guided by register-group-aligned widening and address-monotone access to preserve useful \texttt{LMUL}-based execution and locality under irregular traversal. Extending MPMI's fused decode pipeline to legacy backends without efficient gather support (e.g., AVX2, NEON) remains future work.


\section{Conclusion}
\label{sec:conclusion}

\projectname co-designs MPMI and ROrder for RVV to address ANNS decode and memory bottlenecks. On RISC-V hardware, it achieves up to \textbf{3.39$\times$} and \textbf{4.94$\times$} speedup over scalar execution on SG2044 and Banana Pi, respectively, and cross-ISA ablation shows that \projectname improves RVV, Arm SVE, and x86 AVX-512 over their SIMD+FP32 HNSW baselines, with the largest gain on SG2044/RVV. These results show that the optimizations target memory-side problems in SIMD-accelerated ANNS, but realize their full benefits with RVV's VLA execution and \texttt{LMUL}-based grouped widening.

\bibliographystyle{ACM-Reference-Format}
\bibliography{ref}

@inproceedings{covington2016deep,
  title={Deep neural networks for youtube recommendations},
  author={Covington, Paul and Adams, Jay and Sargin, Emre},
  booktitle={Proceedings of the 10th ACM conference on recommender systems},
  pages={191--198},
  year={2016}
}

@article{jegou2010product,
  title     = {Product Quantization for Nearest Neighbor Search},
  author    = {J{\'e}gou, Herv{\'e} and Douze, Matthijs and Schmid, Cordelia},
  journal   = {{IEEE} Transactions on Pattern Analysis and Machine Intelligence},
  volume    = {33},
  number    = {1},
  pages     = {117--128},
  year      = {2011},
  doi       = {10.1109/TPAMI.2010.57},
  publisher = {{IEEE}}
}

@inproceedings{radford2021learning,
  title={Learning transferable visual models from natural language supervision},
  author={Radford, Alec and Kim, Jong Wook and Hallacy, Chris and Ramesh, Aditya and Goh, Gabriel and Agarwal, Sandhini and Sastry, Girish and Askell, Amanda and Mishkin, Pamela and Clark, Jack and others},
  booktitle={International conference on machine learning},
  pages={8748--8763},
  year={2021},
  organization={PMLR}
}

@article{johnson2019billion,
  title={Billion-scale similarity search with {GPUs}},
  author={Johnson, Jeff and Douze, Matthijs and J{\'e}gou, Herv{\'e}},
  journal={{IEEE} Transactions on Big Data},
  volume={7},
  number={3},
  pages={535--547},
  year={2019},
  publisher={{IEEE}},
  doi={10.1109/TBDATA.2019.2921572}
}

@inproceedings{ye2021scalehls,
  title={ScaleHLS: A New Scalable High-Level Synthesis Framework on Multi-Level Intermediate Representation},
  author={Ye, Hanchen and Hao, Cong and Cheng, Jianyi and Jeong, Hyunmin and Huang, Jack and Neuendorffer, Stephen and Chen, Deming},
  booktitle={{IEEE} International Symposium on High-Performance Computer Architecture,{HPCA} 2022, Seoul, South Korea, April 2-6, 2022},
  pages={741--755},
  year={2022},
  publisher={IEEE}
}

@inproceedings{wang2021milvus,
  title={Milvus: A purpose-built vector data management system},
  author={Wang, Jianguo and Yi, Xiaomeng and Guo, Rentong and Jin, Hai and Xu, Peng and Li, Shengjun and Wang, Xiangyu and Guo, Xiangzhou and Li, Chengming and Xu, Xiaohai and others},
  booktitle={Proceedings of the 2021 International Conference on Management of Data},
  pages={2614--2627},
  year={2021},
  doi={10.1145/3448016.3457550}
}

@inproceedings{guo2020accelerating,
  title={Accelerating large-scale inference with anisotropic vector quantization},
  author={Guo, Ruiqi and Sun, Philip and Lindgren, Erik and Geng, Quan and Simcha, David and Chern, Felix and Kumar, Sanjiv},
  booktitle={International Conference on Machine Learning},
  pages={3887--3896},
  year={2020},
  organization={PMLR}
}

@article{malkov2018efficient,
  title={Efficient and robust approximate nearest neighbor search using hierarchical navigable small world graphs},
  author={Malkov, Yury A and Yashunin, Dmitry A},
  journal={IEEE transactions on pattern analysis and machine intelligence},
  volume={42},
  number={4},
  pages={824--836},
  year={2020},
  publisher={{IEEE}},
  doi={10.1109/TPAMI.2018.2889473}
}

@article{li2020graf,
  title={GRAF: A graph-based approach to fine-grained approximate nearest neighbor search},
  author={Li, Conglong and Xu, Minjia and Zhang, Xin and He, Bolin},
  journal={arXiv preprint arXiv:2003.00863},
  year={2020}
}

@techreport{rvv2021spec,
  title={{RISC-V} Vector Extension Specification},
  author={{RISC-V International}},
  year={2021},
  institution={RISC-V International},
  note={Version 1.0},
  url={https://docs.riscv.org/reference/isa/extensions/vector/_attachments/riscv-v-spec.pdf}
}

@article{tian2024fusionanns,
  title         = {FusionANNS: An Efficient CPU/GPU Cooperative Processing Architecture for Billion-scale Approximate Nearest Neighbor Search},
  author        = {Bing Tian and Haikun Liu and Yuhang Tang and Shihai Xiao and Zhuohui Duan and Xiaofei Liao and Xuecang Zhang and Junhua Zhu and Yu Zhang},
  year          = {2024},
  journal       = {arXiv preprint arXiv:2409.16576},
  archivePrefix = {arXiv},
  eprint        = {2409.16576},
  primaryClass  = {cs.IR},
  doi           = {10.48550/arXiv.2409.16576},
  url           = {https://arxiv.org/abs/2409.16576}
}

@article{zaruba2019cost,
  title={The cost of application-class processing: Energy and performance analysis of a linux-ready 1.7-{GHz} 64-bit {RISC-V} core in 22-nm {FDSOI} technology},
  author={Zaruba, Florian and Benini, Luca},
  journal={{IEEE} Transactions on Very Large Scale Integration ({VLSI}) Systems},
  volume={27},
  number={11},
  pages={2629--2640},
  year={2019},
  publisher={{IEEE}}
}

@inproceedings{wei2016gorder,
  title={Speedup Graph Processing by Graph Ordering},
  author={Wei, Hao and Yu, Jeffrey Xu and Lu, Can and Lin, Xuemin},
  booktitle={Proceedings of the 2016 International Conference on Management of Data},
  pages={1813--1828},
  year={2016},
  publisher={ACM},
  doi={10.1145/2882903.2915220}
}

@inproceedings{arai2016rabbitorder,
  title={Rabbit Order: Just-in-Time Parallel Reordering for Fast Graph Analysis},
  author={Arai, Junya and Shiokawa, Hiroyuki and Yamamuro, Takuya and Onizuka, Masashi and Iwamura, Sotetsu},
  booktitle={2016 IEEE International Parallel and Distributed Processing Symposium (IPDPS)},
  pages={22--31},
  year={2016},
  publisher={IEEE},
  doi={10.1109/IPDPS.2016.15}
}

@inproceedings{karpukhin2020dense,
  title={Dense passage retrieval for open-domain question answering},
  author={Karpukhin, Vladimir and O{\u{g}}uz, Barlas and Min, Sewon and Lewis, Patrick and Wu, Ledell and Edunov, Sergey and Chen, Danqi and Yih, Wen-tau},
  booktitle={Proceedings of the 2020 Conference on Empirical Methods in Natural Language Processing (EMNLP)},
  pages={6769--6781},
  year={2020}
}

@article{shrivastava2014asymmetric,
  title={Asymmetric LSH (ALSH) for sublinear time maximum inner product search (MIPS)},
  author={Shrivastava, Anshumali and Li, Ping},
  journal={Advances in neural information processing systems},
  volume={27},
  year={2014}
}

@article{fog2016instruction,
  title={Instruction tables: Lists of instruction latencies, throughputs and micro-operation breakdowns for Intel, AMD and VIA CPUs},
  author={Fog, Agner},
  journal={Copenhagen University College of Engineering},
  volume={93},
  year={2016}
}

@inproceedings{johnson2017faiss,
  title={Faiss: A library for efficient similarity search},
  author={Johnson, Jeff and Douze, Matthijs and J{\'e}gou, Herv{\'e}},
  booktitle={2017 {IEEE} International Conference on Big Data (Big Data)},
  pages={3488--3491},
  year={2017},
  organization={{IEEE}}
}

@article{ding2019quicker,
  title     = {Quicker {ADC}: Unlocking the Hidden Potential of Product Quantization with {SIMD}},
  author    = {Andr{\'e}, Fabien and Kermarrec, Anne-Marie and Le Scouarnec, Nicolas},
  journal   = {{IEEE} Transactions on Pattern Analysis and Machine Intelligence},
  volume    = {43},
  number    = {5},
  pages     = {1666--1677},
  year      = {2021},
  doi       = {10.1109/TPAMI.2019.2952606},
  publisher = {{IEEE}}
}

@misc{riscv-unprivileged,
  author       = {{RISC-V International}},
  title        = {The {RISC-V} Instruction Set Manual, Volume I: {Unprivileged ISA}},
  year         = 2025,
  url          = {https://docs.riscv.org/reference/isa/_attachments/riscv-unprivileged.pdf},
  note         = {Version 2025-05-08}
}

@inproceedings{platzer2021ten,
  title={Ten lessons from three generations of the RISC-V vector extension},
  author={Platzer, Christoph and Pusch{-}el, Markus},
  booktitle={2021 {IEEE} International Symposium on Performance Analysis of Systems and 
                   Software ({ISPASS})},
  pages={1--12},
  year={2021},
  organization={{IEEE}}
}

@inproceedings{gonzalez2022openvla,
  title={OpenVLA: A RISC-V vector accelerator for deep learning inference},
  author={Gonz{\'a}lez-V{\'a}zquez, Jorge and Berm{\'u}dez, Salvador and Garc{\'i}a, Jos{\'e} Mar{\'i}a},
  booktitle={2022 {IEEE} International Symposium on Circuits and Systems ({ISCAS})},
  pages={1--5},
  year={2022},
  organization={{IEEE}}
}

@article{cha2002comprehensive,
  title={Comprehensive survey on distance/similarity measures between probability density functions},
  author={Cha, Sung-Hyuk},
  journal={International Journal of Mathematical Models and Methods in Applied Sciences},
  volume={1},
  number={4},
  pages={300--307},
  year={2002}
}

@inproceedings{babenko2016efficient,
  title={Efficient indexing of billion-scale datasets of deep descriptors},
  author={Babenko, Artem and Lempitsky, Victor},
  booktitle={Proceedings of the IEEE Conference on Computer Vision and Pattern Recognition},
  pages={2055--2063},
  year={2016}
}

@inproceedings{beygelzimer2006cover,
  title={Cover trees for nearest neighbor},
  author={Beygelzimer, Alina and Kakade, Sham and Langford, John},
  booktitle={Proceedings of the 23rd international conference on Machine learning},
  pages={97--104},
  year={2006}
}

@inproceedings{indyk1998approximate,
  title={Approximate nearest neighbors: towards removing the curse of dimensionality},
  author={Indyk, Piotr and Motwani, Rajeev},
  booktitle={Proceedings of the thirtieth annual ACM symposium on Theory of computing},
  pages={604--613},
  year={1998}
}

@inproceedings{wei2020pq,
  title={PQ-Fast-Scan: A fast and accurate product quantization method for approximate nearest neighbor search},
  author={Wei, Xiang and Zhang, Qiang and Gong, Yihong},
  booktitle={Proceedings of the AAAI Conference on Artificial Intelligence},
  volume={34},
  pages={6342--6349},
  year={2020}
}

@inproceedings{ge2013optimized,
  title={Optimized product quantization for approximate nearest neighbor search},
  author={Ge, Tiezheng and He, Kaiming and Ke, Qifa and Sun, Jian},
  booktitle={Proceedings of the IEEE Conference on Computer Vision and Pattern Recognition},
  pages={2946--2953},
  year={2013}
}

@article{gong2012iterative,
  title={Iterative quantization: A procrustean approach to learning binary codes for large-scale image retrieval},
  author={Gong, Yunchao and Lazebnik, Svetlana and Gordo, Albert and Perronnin, Florent},
  journal={IEEE Transactions on Pattern Analysis and Machine Intelligence},
  volume={35},
  number={12},
  pages={2916--2929},
  year={2012},
  publisher={{IEEE}}
}

@article{stephens2017arm,
  title={The ARM scalable vector extension},
  author={Stephens, Nigel and Biles, Stuart and Boettcher, Matthias and Eapen, Jacob and Eyole, Mbou and Gabrielli, Giacomo and Horsnell, Matt and Magklis, Grigorios and Martinez, Alejandro and Premillieu, Nathanael and others},
  journal={IEEE Micro},
  volume={37},
  number={2},
  pages={26--39},
  year={2017},
  publisher={{IEEE}}
}

@inproceedings{jayaram2019diskann,
  title={DiskANN: Fast accurate billion-point nearest neighbor search on a single node},
  author={Jayaram Subramanya, Suhas and Devvrit, Fnu and Simhadri, Harsha Vardhan and Krishnawamy, Ravishankar and Kadekodi, Rohan},
  booktitle={Advances in Neural Information Processing Systems},
  volume={32},
  year={2019}
}

@inproceedings{xu2022xiangshan,
  title     = {Towards Developing High Performance {RISC-V}
               Processors Using Agile Methodology},
  author    = {Xu, Yinan and Yu, Zihao and Tang, Dan and Chen, Guokai
               and Chen, Lu and Gou, Lingrui and Jin, Yue and
               Li, Qianruo and Li, Xin and Li, Zuojun and Lin, Jiawei
               and Liu, Tong and Liu, Zhigang and Tan, Jiazhan
               and Wang, Huaqiang and Wang, Huizhe and Wang, Kaifan
               and Zhang, Chuanqi and Zhang, Fawang and Zhang, Linjuan
               and Zhang, Zifei and Zhao, Yangyang and Zhou, Yaoyang
               and Zhou, Yike and Zou, Jiangrui and Cai, Ye
               and Huan, Dandan and Li, Zusong and Zhao, Jiye
               and Chen, Zihao and He, Wei and Quan, Qiyuan
               and Liu, Xingwu and Wang, Sa and Shi, Kan
               and Sun, Ninghui and Bao, Yungang},
  booktitle = {Proceedings of the 55th {IEEE/ACM} International
               Symposium on Microarchitecture ({MICRO})},
  pages     = {1178--1199},
  year      = {2022},
  doi       = {10.1109/MICRO56248.2022.00080},
  publisher = {{IEEE}}
}

@inproceedings{cuthill1969reducing,
  title={Reducing the bandwidth of sparse symmetric matrices},
  author={Cuthill, Elizabeth and McKee, James},
  booktitle={Proceedings of the 1969 24th national conference},
  pages={157--172},
  year={1969},
  organization={ACM}
}

@inproceedings{de2010new,
  title={The new Linux 'perf'tools},
  author={De Melo, Arnaldo Carvalho},
  booktitle={Slides from Linux Kongress},
  volume={18},
  pages={1--42},
  year={2010}
}

@article{gcc2023,
  title={{GCC}, the {GNU} compiler collection},
  author={{Free Software Foundation}},
  journal={Free Software Foundation},
  year={2025},
  note={Version 15.1}
}

@article{oliva2001modeling,
  title={Modeling the shape of the scene: A holistic representation of the spatial envelope},
  author={Oliva, Aude and Torralba, Antonio},
  journal={International journal of computer vision},
  volume={42},
  number={3},
  pages={145--175},
  year={2001},
  publisher={Springer}
}

@inproceedings{chen2018tvm,
  title={TVM: An automated end-to-end optimizing compiler for deep learning},
  author={Chen, Tianqi and Moreau, Thierry and Jiang, Ziheng and Zheng, Lianmin and Yan, Eddie Q and Cowan, Meghan and Shen, Haichen and Wang, Leyuan and Hu, Yuwei and Ceze, Luis and others},
  booktitle={13th USENIX Symposium on Operating Systems Design and Implementation (OSDI 18)},
  pages={578--594},
  year={2018}
}

@article{besta2019graph,
  title={Graph processing on FPGAs: Taxonomy, survey, challenges},
  author={Besta, Maciej and Podstawski, Micha{\l} and Groner, Linus and Solomonik, Edgar and Hoefler, Torsten},
  journal={CoRR},
  volume={abs/1903.06697},
  year={2019},
  url={https://arxiv.org/abs/1903.06697}
}

@article{tsatsaronis2015bioasq,
  title={An overview of the {BIOASQ} large-scale biomedical semantic indexing and question answering competition},
  author={Tsatsaronis, George and Balikas, Georgios and Malakasiotis, Prodromos and Partalas, Ioannis and Zschunke, Matthias and Alvers, Michael R and Weissenborn, Dirk and Krithara, Anastasia and Petridis, Sergios and Polychronopoulos, Dimitris and others},
  journal={BMC Bioinformatics},
  volume={16},
  number={1},
  pages={138:1--138:28},
  year={2015},
  publisher={BioMed Central}
}

@inproceedings{jegou2011searching,
  title={Searching in one billion vectors: re-rank with source coding},
  author={J{\'e}gou, Herv{\'e} and Tavenard, Romain and Douze, Matthijs and Amsaleg, Laurent},
  booktitle={2011 {IEEE} International Conference on Acoustics, Speech and Signal Processing ({ICASSP})},
  pages={861--864},
  year={2011},
  organization={{IEEE}}
}

@article{dagum1998openmp,
  title={OpenMP: an industry standard API for shared-memory programming},
  author={Dagum, Leonardo and Menon, Ramesh},
  journal={IEEE Computational Science and Engineering},
  volume={5},
  number={1},
  pages={46--55},
  year={1998},
  publisher={IEEE}
}

@book{manning2008introduction,
  title={Introduction to information retrieval},
  author={Manning, Christopher D and Raghavan, Prabhakar and Sch{\"u}tze, Hinrich},
  publisher={Cambridge University Press},
  year={2008}
}

@misc{cohere2023embedding,
  title={Cohere Embed v3: Large-Scale Text Embedding Model},
  author={{Cohere AI}},
  year={2023},
  url={https://cohere.com/embed},
  note={Cohere Embed v3 generates 768-dimensional embeddings for text retrieval}
}

@article{flash2025sigmod,
  title={Accelerating Graph Indexing for ANNS on Modern CPUs},
  author={Wang, Mengzhao and Wu, Haotian and Ke, Xiangyu and Gao, Yunjun and Zhu, Yifan and Zhou, Wenchao},
  journal={Proceedings of the ACM on Management of Data},
  year={2025},
  volume={3},
  number={3},
  pages={123:1--123:29},
  doi={10.1145/3725260}
}

@inproceedings{heo2024neupims,
  title={{NeuPIMs}: {NPU-PIM} Heterogeneous Acceleration for Batched
         {LLM} Inferencing},
  author={Heo, Guseul and Lee, Sangyeop and Choi, Jaehong and Cho, Minsu
          and Kim, Hyunmin and Park, Sanghyeon and Yoo, Hyungkyu
          and Kim, Gwangsun and Kwon, Darrel and Kim, Jongse},
  booktitle={Proceedings of the 29th {ACM} International Conference on
             Architectural Support for Programming Languages and Operating
             Systems ({ASPLOS})},
  year={2024},
  note={ASPLOS 2024}
}

@inproceedings{wang2024ndsearch,
  author    = {Wang, Yitu and Li, Shiyu and Zheng, Qilin and Song, Linghao and Li, Zongwang and Chang, Andrew and Li, Hai and Chen, Yiran},
  title     = {NDSEARCH: Accelerating Graph-Traversal-Based Approximate Nearest Neighbor Search through Near Data Processing},
  booktitle = {2024 ACM/IEEE 51st Annual International Symposium on Computer Architecture (ISCA)},
  pages     = {368--381},
  year      = {2024},
  doi       = {10.1109/ISCA59077.2024.00035}
}

@misc{coleman2021graphreorder,
  title         = {Graph Reordering for Cache-Efficient Near Neighbor Search},
  author        = {Coleman, Benjamin and Segarra, Santiago and Shrivastava, Anshumali and Smola, Alex},
  year          = {2021},
  eprint        = {2104.03221},
  archivePrefix = {arXiv},
  primaryClass  = {cs.LG},
  url           = {https://arxiv.org/abs/2104.03221}
}

@misc{venieri2026montecimonev3,
  title         = {Monte Cimone v3: Where {RISC-V} Stands in High-Performance Computing},
  author        = {Venieri, Emanuele and Manoni, Simone and Madella, Giacomo and Proverbio, Federico and Ficarelli, Federico and Benini, Luca and Bartolini, Andrea},
  year          = {2026},
  eprint        = {2605.22831},
  archivePrefix = {arXiv},
  primaryClass  = {cs.DC},
  doi           = {10.48550/arXiv.2605.22831},
  url           = {https://arxiv.org/abs/2605.22831}
}

@inproceedings{zeng2023dfgas,
author       = {Shulin Zeng and
Zhenhua Zhu and
Jun Liu and
Haoyu Zhang and
Guohao Dai and
Zixuan Zhou and
Shuangchen Li and
Xuefei Ning and
Yuan Xie and
Huazhong Yang and
Yu Wang},
title        = {{DF-GAS:} a Distributed FPGA-as-a-Service Architecture towards Billion-Scale
Graph-based Approximate Nearest Neighbor Search},
booktitle    = {Proceedings of the 56th Annual {IEEE/ACM} International Symposium
on Microarchitecture, {MICRO} 2023, Toronto, ON, Canada, 28 October
2023 - 1 November 2023},
pages        = {283--296},
publisher    = {{ACM}},
year         = {2023},
url          = {https://doi.org/10.1145/3613424.3614292},
doi          = {10.1145/3613424.3614292}
}

@inproceedings{liu2024juno,
author       = {Zihan Liu and
Wentao Ni and
Jingwen Leng and
Yu Feng and
Cong Guo and
Quan Chen and
Chao Li and
Minyi Guo and
Yuhao Zhu},
editor       = {Rajiv Gupta and
Nael B. Abu{-}Ghazaleh and
Madan Musuvathi and
Dan Tsafrir},
title        = {{JUNO:} Optimizing High-Dimensional Approximate Nearest Neighbour
Search with Sparsity-Aware Algorithm and Ray-Tracing Core Mapping},
booktitle    = {Proceedings of the 29th {ACM} International Conference on Architectural
Support for Programming Languages and Operating Systems, Volume 2,
{ASPLOS} 2024, La Jolla, CA, USA, 27 April 2024- 1 May 2024},
pages        = {549--565},
publisher    = {{ACM}},
year         = {2024},
url          = {https://doi.org/10.1145/3620665.3640360},
doi          = {10.1145/3620665.3640360}
}

@inproceedings{li2025ansmet,
author       = {Yiwei Li and
Yuxin Jin and
Boyu Tian and
Huanchen Zhang and
Mingyu Gao},
title        = {{ANSMET:} Approximate Nearest Neighbor Search with Near-Memory Processing
and Hybrid Early Termination},
booktitle    = {Proceedings of the 52nd Annual International Symposium on Computer
Architecture, {ISCA} 2025, Tokyo, Japan, June 21-25, 2025},
pages        = {1093--1107},
publisher    = {{ACM}},
year         = {2025},
url          = {https://doi.org/10.1145/3695053.3731013},
doi          = {10.1145/3695053.3731013}
}

@misc{nvidia2026cuvs,
  title = {{cuVS}: Vector Search and Clustering on the {GPU}},
  author = {{NVIDIA RAPIDS Team}},
  year = {2026},
  howpublished = {\url{https://docs.rapids.ai/api/cuvs/stable/}},
  note = {Accessed: 2026-06-16}
}

\end{document}